%% file: Third_medium_contact-thermal.tex
\documentclass[12pt,twoside]{article}
\usepackage{a4}

\usepackage[ngerman,english]{babel}

\usepackage{amsmath}
\usepackage{amssymb}
\usepackage{amsbsy}
\usepackage{breqn}
\usepackage{xcolor} 
\usepackage{pgfplots}
\pgfplotsset{compat=1.8}

\usepackage{hyperref}

\usepackage{caption,longtable,booktabs}
\usepackage{subcaption}

\usepackage{epsfig}
\usepackage[round]{natbib}
\usepackage{hyperref}

\usepackage{tikz}

\usepackage{ulem}

\definecolor{myblue}{RGB}{135, 186, 230}
\definecolor{myorange}{RGB}{255, 165, 0}
\definecolor{myred}{RGB}{255, 0, 0}

\newcommand{\atatat}{@}

\usepackage[labelfont=bf, 
format=plain, 
labelsep=endash, 
justification=raggedright 
]{caption}

\input{BH_DEFS}

\input{PWc_DEFS}

\begin{document}
{\Large \bf A third medium approach for thermo-mechanical contact based on low order ansatz spaces}\\ \vspace{10 mm}

\begin{center}
P. Wriggers${}^{1}$  
\end{center}

${}^{1}$  
Leibniz Universit\"at Hannover, Institute of Continuum Mechanics,   Germany,   \\
\hspace{3mm}${}^{1}$ e-mail: wriggers\atatat ikm.uni-hannover.de \\

%
%

%

\vspace{4mm}
\begin{center}
{\bf \large Abstract}
\bigskip
\vspace{1cm}
{\footnotesize
\begin{minipage}{14.5cm}
\noindent
The third medium contact approach has been successfully employed in structural applications and extended to various optimization problems. This discretization technique replaces classical contact formulations and algorithms by introducing a compliant interfacial layer - referred to as the third medium - between the contacting bodies. Unlike traditional contact methods, this formulation naturally accommodates finite deformations at the interface. As the two bodies approach each other, the third medium undergoes compression and effectively acts as a deformable barrier, preventing interpenetration and transmitting contact forces in a smooth and numerically stable manner.
In thermo-mechanical problems, heat conduction must be incorporated into the model, which typically requires specialized interface laws when using classical contact formulations. These laws aim to capture the complex thermal behavior at the contact interface, including discontinuities and varying conductance. In contrast, the third medium approach offers a significant advantage: the thermo-mechanical formulation inherently accounts for the interface behavior without the need for additional interface conditions. This includes the gradual heat transfer through the surrounding gas when the bodies are near each other, as well as the localized heat conduction that occurs upon physical contact. As a result, the third medium naturally captures both non-contact and contact-phase thermal conduction within a unified framework. 
In this paper, we propose a new thermo-mechanical model based on a continuum formulation for finite strains and show by means of examples the behaviour of the associated finite element formulation based on linear ansatz functions.  
\end{minipage}
}
\end{center}

\section{Introduction}
\label{sec:intro}
The numerical treatment of contact mechanics using finite element methods has a long and well-established tradition, reflecting its fundamental importance in a wide range of real-world applications (see, e.g., \cite{Wri06b} and \cite{Lau02}). Various strategies for spatial discretization of contact have been developed, including node-to-node, node-to-segment, and mortar-based formulations, with foundational contributions in each area provided in \cite{ChTu71}, \cite{Hal79}, and \cite{Pus04}, respectively. To address the inherent inequality constraints of contact problems, several algorithmic approaches have been proposed - most notably the penalty method, Lagrange multiplier formulations, barrier functions, and augmented Lagrangian techniques. These methods rely additionally on  robust global and local search algorithms and have been extensively studied and successfully applied in both small and large deformation regimes (cf. \cite{Lau02, Wri06b, Yas13}). Furthermore, contact algorithms have been integrated into adaptive finite element frameworks to accommodate evolving contact conditions and mesh distortion, as demonstrated in \cite{CaScWr99, KiLaDo14}. This need for specialized formulations extends naturally to thermo-mechanical contact problems, where thermal and mechanical fields are coupled at the contact interface. Over the years, several finite element-based approaches have been developed to address this challenging class of problems. For small strain formulations, early contributions include, for example, \cite{WrZa93a}, while significant progress in the context of finite strain thermo-mechanical contact has been made in works such as \cite{JoKl93}, \cite{WrMi94}, \cite{Age98}, and \cite{SeWaPo19}. These formulations incorporate temperature-dependent material behavior and heat transfer across potentially non-smooth, evolving contact interfaces. Furthermore, many of these methods have been successfully extended to adaptive finite element schemes to efficiently handle localized thermal and mechanical gradients, as demonstrated in \cite{RiWr04}.

An alternative approach that circumvents the intricate handling of inequality constraints inherent in classical contact mechanics is offered by the third medium contact method. This method was initially proposed in \cite{WrScSc13}, where an anisotropic constitutive behavior of the third medium near the contact interface was assumed to replicate realistic contact responses. A related concept involves the insertion of a narrow interfacial domain between contacting bodies, as introduced in \cite{OlHaCaWeHe09} and \cite{HaOlWeCaHe09}. However, this variant does not fully embed the interacting bodies within a surrounding medium, thereby still requiring a global contact search strategy. Since its introduction, the third medium method has undergone significant refinement and has found promising applications in areas such as shape optimization and soft robotics. The methodology has also been successfully integrated into isogeometric analysis for contact problems \cite{KrNgWrLo18}, and extended to meshfree coupling techniques \cite{HuNgZh18}. Furthermore, a variant employing alternative constitutive models and higher-order finite elements was developed in \cite{BoZaKoRa15}, enabling the simulation of very large deformations in highly flexible cellular structures.

Recent advances have extended the third medium methodology introduced in \cite{WrScSc13} to optimization-driven applications. In particular, the work in \cite{BlSiPo21} introduced specialized regularization techniques to control excessive element deformation within the third medium, thereby enhancing numerical stability and robustness. This approach was further refined in subsequent studies \cite{BlSiPo23}, \cite{FrSiPo24}, and \cite{FrDaPoSi24}, which addressed the challenges arising from the severe distortions encountered in the discretization of the third medium, especially under large deformations. Beyond optimization, the third medium concept has also been successfully applied to the modeling of metamaterials, as demonstrated in \cite{DaSoWaWaTo23}, where its ability to handle complex interfacial behaviors proved advantageous. Furthermore, the framework has been extended to incorporate frictional contact, employing a novel analogy to crystal plasticity to model interfacial shear behavior, as detailed in \cite{FrRoPoSiGe24}.

Furthermore, the work presented in \cite{FaHoDoRo24} extended the third medium approach to the modeling of pneumatic structures, introducing an innovative regularization technique based on the polar decomposition of the deformation gradient to ensure numerical stability under large deformations. A first step toward incorporating thermo-mechanical behavior within third medium contact formulations is presented in \cite{DaAlFrPoSi25}. This study employs second-order finite elements in combination with the regularization strategy proposed in \cite{FrSiPo24}, and demonstrates the approach through topology optimization of thermo-mechanical regulators, marking a advancement toward fully coupled multiphysics applications.

In the previously cited works, regularization is achieved by incorporating the gradient of the deformation gradient, which necessitates the use of finite elements with at least quadratic shape functions. It is worth noting, however, that quadratic triangular elements are not optimal in this context, as they yield a constant gradient of the deformation gradient $\boldsymbol{F}$, which fails to provide adequate control over element distortion. To address this limitation, a novel regularization technique was proposed in \cite{WrKoJu25} and \cite{WrKoJu25b}, enabling the use of linear quadrilateral and linear triangular finite elements for approximating the gradients of $\boldsymbol{F}$. This advancement significantly broadens the applicability and computational efficiency of regularized third medium formulations.

This paper applies the recently developed technique introduced in \cite{WrKoJu25b} to extend the third medium contact methodology to thermo-mechanical problems. In such problems, heat conduction must be incorporated into the contact model - a task that typically requires specialized interface laws within classical contact formulations, as seen in \cite{ZaWrStSc92a} and \cite{ZaWrStSc92b}. These traditional approaches involve constitutive equations designed to capture the complex thermal behavior at the contact interface, including temperature discontinuities and pressure-dependent thermal conductance. In contrast, the third medium approach presents a distinct advantage: the thermo-mechanical formulation intrinsically captures the interfacial thermal behavior without necessitating additional interface conditions. This includes gradual heat transfer mediated by the surrounding gas when bodies are in close proximity, as well as localized conduction upon actual contact. As such, the third medium framework provides a unified and consistent representation of both non-contact and contact-phase heat transfer, simplifying the modeling process while preserving physical fidelity.

The associated regularization strategy is based on the introduction of an auxiliary scalar field, which enables the computation of the gradient of the deformation gradient $\boldsymbol{F}$. In two-dimensional settings, this approach permits the use of linear triangular and quadrilateral elements, thereby enhancing computational efficiency without sacrificing accuracy. In contrast, the extension to three-dimensional problems requires a modified regularization technique to efficiently approximate the gradients necessary for controlling element distortion - an approach that will be introduced and detailed in this paper and applied to  thermo-mechanical problems..For that, a fully coupled thermo-mechanical formulation for finite deformations will be presented, incorporating pressure-dependent heat transfer across the contact interface. Numerical examples will be provided to illustrate the effectiveness and robustness of the proposed methodology.

\section{General formulation and continuum model}
\label{sec:Continuum_TM}

The approach of two contacting solids $\Omega^{(\alpha)}$ with ($\alpha =\{ 1,2 \} $) is formulated within the third medium methodology  where the third medium $\Omega^{(m)}$ surrounds the solids $\Omega^{(\alpha)}$. Thus, the heat conduction problem is formulated over the entire domain:
$
\Omega = \Omega^{(1)} \cup \Omega^{(2)} \cup \Omega^{(m)}
$. 
The classical interface conditions are no longer applied explicitly at the contact boundary; instead, they are inherently embedded within the interior of the finite element mesh, allowing the thermal and mechanical interactions between contacting bodies to be naturally resolved through the constitutive behavior of the interposed material.

No explicit contact interface conditions are required. Instead, the varying stiffness of the third medium provides a barrier to prevent interpenetration. Obviously, the stiffness of the third medium has to be increased massively to prevent penetration when the two bodies get close to each other. This way of enforcing the contact constraints is actually known as barrier method, see e.g. \cite{Wri06b}, however here formulated using continuum elements. Furthermore, heat flux through the third medium governs thermal interaction between $\Omega^{(1)}$ and $\Omega^{(2)}$. When the solids are separated, conduction occurs primarily through gas (low conductivity coefficient). Upon contact, the conductivity coefficient may be significantly increased to model solid-solid conduction.

\begin{figure}[h]%
	\centering
	\includegraphics[width=1.0\textwidth]{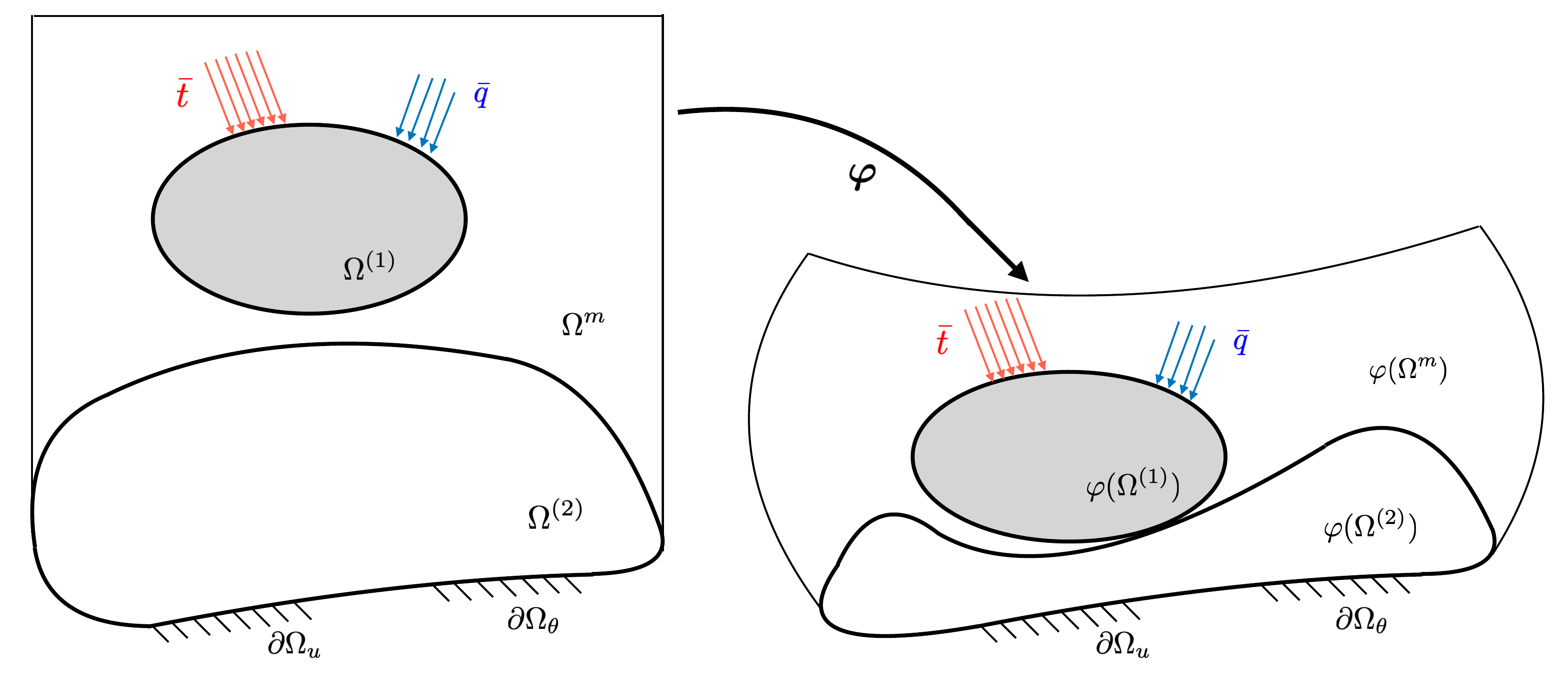}
	\caption{General idea of third medium contact: the third medium $\Omega^{m}$ is placed around the two physical bodies $\Omega^{(1)}$ and $\Omega^{(2)}$ (left). During the loading the solids undergo the deformation $\boldsymbol{\varphi}$ and the two bodies come into contact. The third medium is severely deformed (right).}
	\label{fig:Solids_TM}
\end{figure}

Based on these modeling assumptions if is clear that the third medium will be highly compressed in contact. The general idea is depicted in Fig. \ref{fig:Solids_TM} where $\varphi( \Omega^{(\alpha)})$ with $\alpha =\{ 1,2,m\} $ denotes the deformed configuration of the solids and the third medium. 

\subsection{Kinematics}
We consider  solids that may undergo finite deformations upon contact. All governing equations are formulated with respect to the initial configuration. For sake of clarity,  superscripcts distinguishing  the individual solids and the third medium will be omitted unless necessary. 

Let  $\boldsymbol X$  denote a material point in the initial configuration and $t$ the time. The motion of the body is described by hte mapping  $\boldsymbol x = \boldsymbol {\varphi}(\boldsymbol X,t)$ where $\boldsymbol x$ is the position in the current configuration. The associated deformation gradient is given by $\boldsymbol F = \mbox{Grad}\, \boldsymbol {\varphi}$ which can multiplicatively decomposed into a  volumetric and an isochoric part: $J =\det \boldsymbol F$,  $ \widehat{\boldsymbol F}= J ^{-\frac{1}{3}} \boldsymbol F$. Here, $J$
denotes the Jacobi determinant $J =\det \boldsymbol F$, representing local volume change.  

The right Cauchy-Green deformation tensor $\boldsymbol {C}=\boldsymbol {F}^T\boldsymbol {F}$ measures the finite strains. Its   isochoric part is defined as $ \widehat{\boldsymbol C} = J ^{-\frac{2}{3}} \boldsymbol C$. This decomposition facilitates the use of constitutive models that separately account for volumetric and distortional effects, which is particularly important in thermomechanically coupled materials.

In a thermo-meachanical formulation the absolute temperature $\theta(\boldsymbol X\,,t) >0$, given in the initial configuration, influences only the volumetric deformation. By using the multiplicative split of the deformation gradient $ \boldsymbol F=  \boldsymbol F_e \boldsymbol F_\theta$, see \cite{LuPi75}, which leads for    the volumetric part of the deformation gradient to
\beq
J= J_e\,J_\theta
\label{eq:el_vol_strain_solid}
\eeq
where $J_e$ is related to the elastic, mechanical deformation and $J_\theta$ to the thermal volume change.

\subsection{Balance equations and weak forms}
The  balance of linear momentum and the energy equation for stationary heat conduction are given by
\beq
\begin{aligned}
\mbox{Div}\,\boldsymbol P + \rho_0\,\bar {\boldsymbol b} &=\boldsymbol 0\\
- \mbox{Div}\,\boldsymbol Q + \rho_0\,\bar R &=0
\end{aligned}
\label{eq:strong_form}
\eeq
where $\boldsymbol P $ is the $1^{st}$ Piola-Kirchhoff stress tensor, $\bar {\boldsymbol b} $ the body force,  $\boldsymbol Q $ the heat flux vector in the initial configuration, $\rho_0$ the density in the initial configuration and $\bar R$ a heat source. The Dirichlet and Neumann boundary conditions can be formulated for the thermo-elastic solid with respect to the initial configuration as 
\beq
\begin{aligned}
\boldsymbol u= \bar {\boldsymbol u} \quad\mbox{on}\,\,\, \Gamma_u \qquad& \boldsymbol P\,\boldsymbol N= \bar {\boldsymbol t} \quad\mbox{on}\,\,\, \Gamma_t\\
\theta = \bar \theta \quad\mbox{on}\,\,\, \Gamma_T \qquad& \boldsymbol Q\,\boldsymbol N= \bar q  \quad\mbox{on}\,\,\, \Gamma_h
\end{aligned}
\label{eq:D_N_bcs}
\eeq

The associated weak form to \eqref{eq:strong_form} and \eqref{eq:D_N_bcs} consists of the mechanical and the thermal part
\beq
G(\boldsymbol u\,,\theta\,,\delta\boldsymbol u\,,\delta\theta) = G_M(\boldsymbol u\,,\theta\,,\delta\boldsymbol u) +G_\theta(\boldsymbol u\,,\theta\,,\delta\theta) =0
\eeq
with the mechanical part
\beq
G_M(\boldsymbol u\,,\theta\,,\delta\boldsymbol u) = \int_\Omega [\boldsymbol P(\boldsymbol u\,,\theta) \cdot \delta \boldsymbol F - \bar {\boldsymbol b} \cdot \delta \boldsymbol u]\,\mbox{d}\Omega - \int_{\Gamma_t} \,\bar {\boldsymbol t} \cdot \delta \boldsymbol u\,\mbox{d}\Gamma
\label{eq:wf_mech}
\eeq
and the thermal part
\beq
G_\theta(\boldsymbol u\,,\theta\,,\delta\theta) = \int_\Omega [-\boldsymbol Q(\boldsymbol u\,,\theta) \cdot \nabla_X \delta \theta -\bar R \, \delta\theta]\,\mbox{d}\Omega -  \int_{\Gamma_h} \,\bar q\, \delta \theta\,\mbox{d}\Gamma
\label{eq:wf_therm}
\eeq
Here, $\delta \boldsymbol{u}$ denotes the virtual (or admissible) variation of the displacement field, $\delta \boldsymbol{F} = \operatorname{Grad}, \delta \boldsymbol{u}$ is the corresponding variation of the deformation gradient, and $\delta \theta$ represents the variation of the temperature field.

It is important to emphasize that the mechanical and thermal weak forms are strongly coupled through both the displacement and temperature fields. Specifically, in the mechanical weak form \eqref{eq:wf_mech}, the $1^{st}$  Piola-Kirchhoff stress tensor $\boldsymbol{P}$ is a function of both the deformation and temperature, reflecting thermo-mechanical coupling in the constitutive model. Similarly, in the thermal weak form \eqref{eq:wf_therm}, the heat flux vector $\boldsymbol{Q}$ depends on the deformation and temperature, capturing the influence of kinematic changes - such as large deformations - on thermal conduction.

\subsection{Constitutive equations}

In the context of finite deformation formulations for thermo-mechanical problems, the thermal and mechanical fields are intrinsically coupled through both the deformation and temperature dependencies. The thermal response is not only influenced by the evolving temperature field but also significantly affected by the deformation of the solid, which alters its geometry and, consequently, its heat conduction behavior. This coupling is particularly pronounced in large deformation regimes, where changes in the material configuration due to mechanical loading must be accurately accounted for in the heat conduction equation. As a result, the governing equations must consistently reflect the interplay between kinematics and thermal transport to ensure a physically sound and robust modeling framework.

By employing a volumetric-isochoric decomposition of the deformation, a thermodynamically consistent strain energy function $\Psi(\boldsymbol{C}, \theta)$ can be formulated for thermo-elastic materials. This decomposition facilitates the separate treatment of volumetric and deviatoric effects, which is particularly useful in modeling compressible and incompressible behaviors, as well as anisotropic materials -- see, for instance, \cite{HaNe03}, \cite{Sch09}, and \cite{RiWr23}. In this work, we adopt a simplified yet robust Neo-Hookean-type strain energy density function that incorporates thermal effects, as proposed in \cite{GoCa24}, and builds on the volumetric formulation of \cite{SiTa82}:
\beq
\Psi(\boldsymbol C\,,\theta)=  \frac{K}{2} \left [ \ln J \right ]^2 + \frac{\mu}{2}\left ( J^{-\frac{2}{3}} \,\mbox{tr}\,{\boldsymbol{C}} -3 \right ) - 3 \,\alpha_t \,(\theta-\theta_0) \, K\,{\ln J} +c_t\left [ \theta- \theta_0-\theta \ln \frac{\theta}{\theta_0} \right ]  \,.
\label{eq:Psi_th}
\eeq
Here, K and $\mu$ denote the bulk and shear moduli, respectively. The temperature $\theta$ represents the current thermal state, while $\theta_0$ is a reference temperature. The coefficient $\alpha_t$ models thermal expansion, and $c_t$ characterizes the specific heat capacity of the material. The energy function reflects both mechanical and thermal contributions, including a volumetric thermal term and an entropic contribution consistent with classical thermoelasticity.

 The $2^{nd}$ Piola-Kirchhoff stress is then given by
\beq
\boldsymbol S  =2\, \frac{\partial \Psi(\boldsymbol C)}{\partial \boldsymbol C} \qquad \mbox{with} \,\,\,\boldsymbol P =\boldsymbol F\,  \boldsymbol S \,.
\eeq

Note that the thermal volumetric expansion related to the above model is given by, see e.g. \cite{HoSi96},
\beq
J_\theta = e^{3\,\alpha_t\,(\theta-\theta_0)}\,.
\label{eq:thermal_vol_strain_solid}
\eeq

Since this work focuses exclusively on stationary thermo-mechanical problems, the entropic contribution associated with the heat capacity can be omitted from the strain energy function in \eqref{eq:Psi_th}. Accordingly, the simplified strain energy reflects only the elastic and thermal expansion effects relevant to steady-state analyses. The same functional form is also employed to model the constitutive behavior of the third medium. However, in this context, the material parameters -- such as stiffness and thermal conductivity -- are significantly reduced to reflect the physical characteristics of a surrounding gas. This allows the third medium to act as a compliant, thermally conductive interface that facilitates a smooth transition between non-contact and contact states, while capturing essential thermal effects without the need for explicit interface conditions.

The heat flux has to be formulated in the current configuration. It depends on the gradient of the temperature and known as Fourier law. The heat flux for isotropic heat conducation is given by $\boldsymbol q=  - k_\theta\,\nabla_x \theta$  where $ \nabla_x$ denotes the gradient with respect to the coordinates  $\boldsymbol x$  of the current configuration. It can be transformed to the initial configuration using Nanson's formula and $\nabla_x \theta= \boldsymbol F^{-T} \,\nabla_X \theta $  leading to 
\beq
\boldsymbol Q=  -J\, k_\theta\,\boldsymbol F^{-1}\,\boldsymbol F^{-T} \,\nabla_X \theta = -J\, k_\theta\,\boldsymbol C^{-1}\,\nabla_X \theta 
\label{eq:Fourier_law}
\eeq
where $\boldsymbol Q$ is the heat flux with respect to the initial configuration and $k_\theta$ is the thermal conductivity. The gradient operator $\nabla_X \theta $ is computed with respect to the coordinates $\boldsymbol X$ of the initial configuration.

\subsection{Pseudo potential for the thermo-mechanical problem}

From a computational perspective, especially when employing automatic differentiation frameworks such as AceGen (see \cite{KoWr16}), it is highly advantageous to formulate the problem using potential-based expressions. These forms not only ensure consistency in the derivation of governing equations but also lead to more efficient and robust code generation. Following the approach proposed in \cite{WrAlHu23}, a pseudo-potential formulation can be introduced, replacing the classical weak forms and serving as the foundation for the finite element discretization. The thermo-mechanical response of the solid is then governed by the  (pseudo) potential 
\beq
\Pi( \boldsymbol {u}\,,\theta) = \int\limits_{\Omega} \left [ \Psi(\boldsymbol C\,,\theta) - \bar {\boldsymbol{b}} \cdot  \boldsymbol {u} - \bar R\,\theta \,\right ] \,\mbox{d}\Omega - 
\int\limits_{\partial \Omega_\sigma} \bar {\boldsymbol {t}} \cdot  \boldsymbol {u} \,\mbox{d}\Gamma 
-\int\limits_{\partial \Omega_h} \bar {{h}} \cdot  \theta\,\mbox{d}\Gamma \Rightarrow STAT \,,
\label{eq:Potential_solid}
\eeq
allowing for a consistent and efficient numerical implementation of the coupled problem.
Here, $\bar {\boldsymbol{b}} $ and $ \bar {\boldsymbol {t}}$ are the applied body forces and surface tractions, respectively. In this work, we assume that neither the volumetric loading term $\bar {\boldsymbol{b}} $  nor the surface traction  $\bar {\boldsymbol {t}}$ depend on the temperature. The thermo-mechanical pseudo strain energy is given by
\beq
\begin{aligned}
\Psi(\boldsymbol C\,,\theta) &= \Psi_{vol} (J) +\Psi_{iso} (\boldsymbol{C})+ \Psi_{e,th} (J\,,\theta) + \Psi_{th} (\boldsymbol{u}\,,\theta)\\
&=\frac{K}{2} \left [ \ln J \right ]^2 + \frac{\mu}{2}\left ( J^{-\frac{2}{3}} \,\mbox{tr}\,{\boldsymbol{C}} -3 \right ) - \left . 3 \,\alpha_t \,(\theta-\theta_0) \right |_{c} \, K\,{\ln J} - \left . \boldsymbol Q(\boldsymbol{u}\,,\theta)\,\right |_{c} \cdot \nabla_X \theta
\end{aligned}
\label{eq:strain_E_thermo}
\eeq
where $\boldsymbol Q$ is given in \eqref{eq:Fourier_law}. All terms denoted by $\left . () \right |_{c}$ have to be kept constant during the first variation of $\Psi$ in order to obtain the correct weak forms in \eqref{eq:wf_mech} and \eqref{eq:wf_therm}. The thermo-mechanical coupling in the formulation arises primarily from two sources: the thermal expansion contribution, denoted by $\Psi_{e,th}$ in the strain energy function, and the dependence of the heat flux on the deformation, as described in \eqref{eq:Fourier_law}. Additional coupling effects could, in principle, stem from the temperature dependence of the bulk modulus $K$ and the shear modulus $\mu$. However, for the sake of clarity and to focus on the use of the third medium approach for thermo-mechanical contact, this temperature dependence of the material parameters is neglected in the present work.

\subsection{Strain energies for thermo-mechanical behaviour of the third medium}
\label{sec:strain_energy_TMC}
The third medium  occupies the space between two contacting bodies acting as a buffer or barrier for the deformation on one side and as a gas that conducts temperature between the solids.  
Its strain energy can be chosen in an arbitrary way. However, its thermal part should obey the physical laws of thermodynamics. 

There exist several requirements for a strain energy function $\Psi_m(\boldsymbol C\,,\theta)$ that models the thermo-mechanical behavior of the third medium: 
\begin{enumerate}
\item $\Psi_m(\boldsymbol C\,,\theta)$ should have almost no influence on the deformation when there is no contact and thus it has to be scaled by a small parameter $\gamma$. 
\item  $\Psi_m(\boldsymbol C\,,\theta)$ has to assume a very high value when the bodies getting close to each other. 
\item  $\Psi_m(\boldsymbol C\,,\theta)$ should be able to represent the physical behaviour of a gas between the solids.
\item  $\Psi_m(\boldsymbol C\,,\theta)$ has to represent the correct conductance between the solids once the bodies are very close.
\end{enumerate}
To achieve this goal, the elastic part of the strain energy $\Psi_{m}(\boldsymbol C\,,\theta)$, for the third body consists of a volumetric  and an isochoric part
\beq
\Psi_{m}(\boldsymbol C\,,\theta)=\frac{\gamma}{2}  \left [\left [ \ln J \right ]^2 + \left ( J^{-\frac{2}{3}} \,\mbox{tr}\,{\boldsymbol{C}} -3 \right )\right ]- \left . 3 \,\alpha_t \,(\theta-\theta_0) \right |_{c} \, \gamma\,{\ln J} - \left . \boldsymbol Q(\boldsymbol{u}\,,\theta)\,\right |_{c} \cdot \nabla \theta \,.
\label{eq:W_medium}
\eeq
The term $ \ln J$ goes for a large compression of the third medium to infinity: $J \rightarrow 0: \Psi_m \rightarrow \infty$.  Due to that, the stiffness of the third medium will be  large  when a solid gets close to another one. In this case, the third medium acts like a barrier function, known from contact mechanics, and avoids penetration. 

The constitutive behaviour of the thermo-elastic response of the third medium is described by the terms $\Psi_{e,th} (J\,,\theta)$ and $ \Psi_{th} (\boldsymbol{u}\,,\theta)$.  Contrary to the formulation of the mechanical part of the third medium, the heat conduction is given a physical meaning of a gas that acts like air between the contacting bodies. This allows a more accurate description of the real thermo-mechanical behaviour and provides a heat conduction through the third medium, even if there is no contact. We note that the conductivity has to be changed to a pressure depending conductivity in the contact interface which can be deduced form micro-mechanical observations related to the roughness of the contacting surfaces, see e.g. \cite{ZaWrStSc92a} and \cite{WrZa93a}. Thus, we can write 
\beq
\boldsymbol Q = - J\, k_{TM}(J)\,\boldsymbol F^{-1}\,\boldsymbol F^{-T} \,\nabla_X \theta =  J\, k_{TM}(J)\,\boldsymbol C^{-1}\,\nabla_X \theta 
\label{eq:Fourier_law_TMC}
\eeq
where $k_{TM}$ is the heat conduction coefficient of the third medium. The heat conduction coefficient $k_{TM}$ of the third medium will be limited to the  heat conduction coefficient of the solids to model real thermo-mechanical contact which is related to the direct contact at the contacting interfaces. This behaviour is modeled by
\beq
\boldsymbol Q= J\, \min ( k_{TM} [\ln J]^2\,,k_\theta) \,\boldsymbol C^{-1}\,\nabla_X \theta 
\eeq
which leads to an increase of the thermal conductivity from $ k_{TM} $ to $k_\theta$ during the approach of two solids and limits the conductivity to $k_\theta$.

\subsection{Regularization of the deformation in the third medium}

When using the strain energies in \eqref{eq:W_medium} and \eqref{eq:strain_E_thermo}, then, the finite elements of the third medium will deform severely in the contacting zone, especially when the contact problem has free boundaries, see e.g. \cite{BlSiPo21}, \cite{FrSiPo24} and \cite{WrKoJu25}.

To ensure numerical stability and prevent excessive mesh distortion in the third medium, it is essential to introduce a suitable regularization mechanism; see, for example, \cite{FrSiPo24}, \cite{FaHoDoRo24}, \cite{FrDaPoSi24}, and \cite{WrKoJu25}. These contributions have resulted in a range of regularization strategies tailored to control the element shape and deformation within the third medium. The formulations presented in \cite{FrSiPo24} and \cite{FaHoDoRo24} rely on higher-order approximations and thus require the use of quadratic finite elements. In contrast, the method proposed in \cite{WrKoJu25} enables the use of both triangular and quadrilateral elements with linear shape functions, significantly reducing computational complexity while maintaining robustness. In the present work, we adopt the latter regularization approach due to its efficiency and broader applicability in large-scale simulations.

Following \cite{WrKoJu25b} the regularization energy is introduced in the two-dimensional case as 
\beq
\Psi_{R\,2}  
=
 \frac{\gamma}{2}  \,\alpha_r\, \|\nabla f^{R}_{1}  \|^2  \quad\mbox{with}\,\,\, f^{R}_{1}=\left [\frac{F_{12}-F_{21}}{F_{11}+F_{22}} \right ] =\tan \varphi_1
\label{eq:W_regular_all_red3}
\eeq 
which contains the components of the deformation gradient.
Note that the numerator in the expression above corresponds to the skew-symmetric (shear) component of the deformation gradient, whereas the denominator captures its symmetric (stretch) part. This regularization approach stems from the idea of utilizing the gradient of the rotation tensor $\boldsymbol{R}$ obtained via the polar decomposition $\boldsymbol{F} = \boldsymbol{R}\, \boldsymbol{U}$. This concept was originally introduced in \cite{FaHoDoRo24}, where a different regularization functional -- distinct from the one considered in \eqref{eq:W_regular_all_red3} -- was employed. The formulation adopted here refines that initial concept by directly targeting mesh distortion through a scalar measure derived from $\|\nabla \boldsymbol{R}\|$ which leads in the two-dimensional case to $\| \nabla \varphi_1 \|$, see \cite{WrKoJu25}. The rotation angle $\varphi_1$ was then replaced by its tangent leading to \eqref{eq:W_regular_all_red3}
thereby offering improved control over element quality in the presence of large deformations.

The approach in \eqref{eq:W_regular_all_red3}  can be extended to the three-dimensional case by assuming that the rotations about the three axes in space are not coupled, leading with
\beq
 f^{R}_{2}=\left [\frac{F_{13}-F_{31}}{F_{33}+F_{11}} \right ] \quad \mbox{and} \,\,\,f^{R}_{3}=\left [\frac{F_{23}-F_{32}}{F_{33}+F_{22}} \right ] 
 \label{eq:def_f_skew}
 \eeq
 and the definition of $ f^{R}_{1}$ in \eqref{eq:W_regular_all_red3} to
\beq
\Psi_{R\,3} =  \frac{\gamma}{2}  \,\alpha_r\, \left (\,  \|\nabla  f^{R}_{1}  \|^2  + \|  \nabla f^{R}_{2}  \|^2 + \| \nabla f^{R}_{3}  \|^2 \,\right ) \,.
\label{eq:W_reg_Hu_skew}
\eeq
This formulation introduces only three components to approximate the rotation gradient of $\nabla \boldsymbol R$ which results in a gradient using a scaled skew symmetric part  of $\boldsymbol F$ and will be advantageous when using the reformulation of the regularization in Section \ref{sec:reduced_grad}. 

All of the aforementioned regularization formulations necessitate at least quadratic interpolation functions, because lower-order elements (e.g., linear quadrilateral Q1 or triangular T1 elements) cannot directly compute second-order gradients such as $\nabla \boldsymbol{F} = \nabla\nabla \boldsymbol{u}$. These second derivatives vanish for linear shape functions, making them unsuitable for evaluating curvature-dependent regularization terms. 
Therefore, classical quadrilateral elements with quadratic shape functions - namely Q2 and Q2S in two dimensions, and H2/H2S in three dimensions - are required for accurately modeling the third medium. These second-order elements incorporate mid-side nodes and enable the interpolation necessary to resolve second derivatives and bending behavior.

 It is important to recognize a critical limitation: in the case of quadratic triangular elements, the gradient of the deformation gradient remains constant over the element domain. This leads to a significant drawback, parti\-cularly in problems involving open  boundaries, where strong mesh distortion of the third medium may occur. Due to the lack of gradient variation within each element, the regularization fails to adequately control the shape of the triangular element, thus compromising its robustness in complex contact scenarios.

In the work by \cite{WrKoJu25}, a significant advancement was introduced that enables the use of finite elements with linear shape functions for the discretization of the third medium. This paper builds upon that methodology by extending it to two-dimensional triangular elements as well as to three-dimensional elements. The resulting formulation provides a robust and computationally efficient framework for the third medium approach which will be discussed in the next section.

\subsection{Approximate computation of the gradients}
\label{sec:reduced_grad}
The direct computation of the gradient of the deformation gradient $\boldsymbol F$, the rotation angle $\varphi$ and the term $\tan\varphi$  is complex and leads to finite elements which are not very efficient at element level. A remedy is to introduce a formulation that approximates these gradients, see \cite{DiHa08} and \cite{JuHa15}.  For a third medium with the strain energy $\Psi_m$ in \eqref{eq:W_medium} and the regularizations in \eqref{eq:W_regular_all_red3} and \eqref{eq:W_reg_Hu_skew} the approach yields 
 in the three-dimensional case, where 
 the regularization term in \eqref{eq:W_reg_Hu_skew} was split into the three components which define the rotation around the different axes. 
 This leads to  
\beq
\begin{aligned}
\Psi^{TMC}_{3d}(\boldsymbol u\,,p_i)&= \Psi_m(\boldsymbol u )+\Psi_{f}(\boldsymbol u, p_i )
\qquad \mbox{with}\\
\Psi_{f}(\boldsymbol u, p_i ) &= \frac{\gamma}{2} \sum_{i=1}^3  \left ( \alpha_1\, \left [f^{R}_{i} -\frac{1}{d} \,p_i \right ]^2+\alpha_2\,  \| \nabla p_i\, \|^2 \right )
\end{aligned}
\label{eq:reg_TMC_3d}
\eeq 
with three additional field variables $p_i$ which again have to be discretized and introduced as additional degrees of freedom in the third medium element.
Equation \eqref{eq:reg_TMC_3d} holds also for the two-dimensional case by only using the component $ f^{R}_{1}$ and the associated additional field $p_1$.

The parameters $\alpha_1$ and $\alpha_2$ must be selected to ensure that the approximation of the deformation gradient is effectively enforced by the regularization functional  \eqref{eq:reg_TMC_3d} which  consists of two principal terms. The first term acts as a penalty that constrains the regularization function $f^{R}_{i} $ to match the auxiliary variable $p_i$, and thus requires a sufficiently large penalty parameter $\alpha_1$ to strongly enforce this constraint. As discussed in \cite{WrKoJu25b}, numerical investigations indicate that the product of $\alpha_1$ with the scaling factor $\gamma$ can be set to $\beta_1 = \alpha_1 \gamma=1$.

The second parameter, $\alpha_2$, governs the influence of the gradient regularization on the overall mesh deformation. To avoid excessive stiffening of the mesh and preserve the physical fidelity of the third medium, this parameter must be chosen sufficiently small. Based on extensive numerical studies in  \cite{WrKoJu25b} the range for the product $\alpha_2\,\gamma$ is $\alpha_2,\gamma = \beta_2 \in [10^{-4},..., 10^{-3}]$. Accordingly, the second part of equation~\eqref{eq:reg_TMC_3d} can be reformulated as:
\beq
 \Psi_{Rf}(\boldsymbol u, p_i ) =  \sum_{i=1}^3  \Psi_{f\,i}(\boldsymbol u, p_i ) =  \sum_{i=1}^3  \left ( \frac{\beta_1}{2} \left [f^{R}_{i} -\frac{1}{d} \,p_i \right ]^2+\frac{\beta_2}{2}\,  \| \nabla p_i\, \|^2  \right )\,.
\label{eq:reg_TMC_3d_beta}
\eeq
This equation  holds also for the two-dimensional case when only the first component is used: $ \Psi_{f\,1}(\boldsymbol u, p_1) $.

The parameter d serves as a global scaling factor that reflects the characteristic size of the computational domain. It is determined as the maximum spatial extent of the structure across all coordinate directions, based on the nodal positions in the undeformed configuration. Specifically, it is computed as
\beq
d= \max ( \max_i X_i -\min_i X_i\,,\max_i Y_i -\min_i Y_i\,,\max_i Z_i -\min_i Z_i)\,.
\eeq
where i ranges over all nodes in the finite element mesh. This scaling ensures that the regularization parameters are independent of the absolute size of the geometry. In the numerical examples presented in this work, the parameters are chosen as $\beta_1 = 1$ and $\beta_2 = 10^{-3}$.

The regularization formulation \eqref{eq:reg_TMC_3d} offers several notable advantages. Foremost among these is its compatibility with both triangular and quadrilateral finite elements using linear ansatz functions in two dimensions, as well as with tetrahedral and hexahedral elements in three-dimensional settings. This flexibility facilitates the use of efficient meshing strategies and significantly lowers computational overhead, particularly when compared to methods that require higher-order elements.

Moreover, these formulations circumvent the need for explicitly computing the gradient terms $ \nabla  f^{R}_i $ and their consistent linearizations, which would otherwise complicate the implementation of the Newton-Raphson solution scheme for the nonlinear system. Instead, the required gradient fields associated with the auxiliary interpolation functions $p_1$, $p_2$ and $p_3$ in \eqref{eq:reg_TMC_3d} can be evaluated in a manner analogous to the standard displacement gradient, thus preserving algorithmic simplicity and numerical robustness.

Naturally, this formulation introduces the drawback of additional degrees of freedom: namely, the inclusion of an auxiliary scalar field $p_1$ in the two-dimensional case, and three scalar fields $p_1$, $p_2$ and $p_3$ in the three-dimensional setting. These fields augment the total number of unknowns in the third medium formulation. However, this increase in computational complexity can be mitigated through various strategies, such as selective condensation, or hierarchical solution techniques, as discussed in detail in \cite{WrKoJu25b}.

Finally, it is important to note that no additional regularization is required for the thermal response of the third medium, as the associated mesh deformation is already effectively governed by the mechanical component of the formulation.
 
 \section{Finite element formulation}
 The finite element disscretization will be derived for elements with linear and quadratic ansatz functions using the classical isoparametric concept. The formulation will be presented for the solids and the third medium. 
 
 \subsection{Finite elements for the solids}
 The discretization for the thermo-elastic solids is based on \eqref{eq:Potential_solid} and \eqref{eq:strain_E_thermo}. In both forms the displacement field $\boldsymbol u$ and the temperature $\theta$ of the solid will  be approximated by the fields $\boldsymbol u_h$ and $\theta_h$ leading to the psudo-potential
 \beq
\Pi( \boldsymbol {u}_h\,,\theta_h) = \int\limits_{\Omega} \left [ \Psi(\boldsymbol C_h\,,\theta_h) - \bar {\boldsymbol{b}} \cdot  \boldsymbol {u}_h - \bar R\,\theta \,\right ] \,\mbox{d}\Omega - 
\int\limits_{\partial \Omega_\sigma} \bar {\boldsymbol {t}} \cdot  \boldsymbol {u}_h \,\mbox{d}\Gamma 
-\int\limits_{\partial \Omega_h} \bar {{h}} \cdot  \theta_h\,\mbox{d}\Gamma \Rightarrow STAT 
\label{eq:Potential_solid_FEM}
\eeq
and
\beq
\begin{aligned}
\Psi&(\boldsymbol C_h\,,\theta_h) = \Psi_{vol} (J_h) +\Psi_{iso} (\boldsymbol{C_h})+ \Psi_{e,th} (J_h\,,\theta_h) + \Psi_{th} (\boldsymbol{u}_h\,,\theta_h)\\
&=\frac{K}{2} \left [ \ln J_h \right ]^2 + \frac{\mu}{2}\left ( J_h^{-\frac{2}{3}} \,\mbox{tr}\,{\boldsymbol{C_h}} -3 \right ) - \left . 3 \,\alpha_t \,(\theta_h-\theta_0) \right |_{c} \, K\,{\ln J_h} - \left . \boldsymbol Q(\boldsymbol{u}_h\,,\theta_h)\,\right |_{c} \cdot \nabla \theta_h
\end{aligned}
\label{eq:strain_E_thermo_FEM}
\eeq

The displacement field $\boldsymbol{u}_h$ and the temperature field $\theta_h$ are approximated in the two- and three-dimensional case using ansatz functions $N_I$, which are associated with the specific finite element type chosen for the solid, see e.g. \cite{ZiTa00b} and \cite{Wri06a}. In general, the finite element approximation is defined by
 \beq
 \boldsymbol{u}_h = \sum_{I=1}^{n}\, N_I(\boldsymbol{\xi}) \,\mathbf{u}_I  \qquad\mbox{and}\,\,\, \theta_h= \sum_{I=1}^{n}\, N_I(\boldsymbol{\xi}) \,\theta_I 
 \label{eq:discr_u_theta}
 \eeq
 where $N_I(\boldsymbol{\xi})$ are the shape function defined with respect to the isoparametric parameter space $\boldsymbol{\xi}$,
 $ \mathbf{u}_I$ and $\theta_I$are the nodal degrees of freedom at node $I$, and $n$ relates to the number of nodes. 
 
   The finite element approximation of the right Cauchy-Green tensor  in  \eqref{eq:Potential_solid_FEM} is then given by $\boldsymbol{C}_h= \boldsymbol{F}^T_h\boldsymbol{F}_h$ where the deformation gradient  $\boldsymbol{F}_h$  follows from
 \beq
 \boldsymbol{F}_h = \boldsymbol{I}+ \nabla_X   \boldsymbol{u}_h =  \boldsymbol{I}+\sum_I \mathbf{u}_I \otimes \nabla_X N_I \,.
  \label{eq:F_h}
 \eeq
Furthermore, the approximation of the Jacobian $J_h$ is computed as $J_h= \det \boldsymbol{F}_h $. In the same way one can compute the temperature gradient
\beq
\nabla \theta_h =  \sum_I \nabla_X N_I \,  \theta_I \,.
  \label{eq:grad_q_h}
 \eeq
By substituting these expressions into \eqref{eq:Potential_solid_FEM} and incorporating the thermo-mechanical strain contributions from \eqref{eq:strain_E_thermo_FEM}, the mechanical and thermal residuals corresponding to a finite element $\Omega_e$ are obtained as
 \beq
 \mathbf {R}^u_e = \frac{\partial \Pi( \boldsymbol {u}_h\,,\theta_h) }{\partial \mathbf u_e} \quad \mbox{and}\,\,\, 
  \mathbf {R}^\theta_e = \frac{\partial \Pi( \boldsymbol {u}_h\,,\theta_h) }{\partial \boldsymbol{\theta}_e}
    \label{eq:Res_u_T}
 \eeq
 where $\mathbf u_e= \{  \mathbf u_1\,,  \mathbf u_2\,, \ldots,  \mathbf u_n\}^T$ and $\boldsymbol{\theta}_e= \{ \theta_1\,,  \theta_2\,, \ldots,  \theta_n\}^T$ are the vectors of the nodal displacements and temperature of the element $\Omega_e$. Note that some parts of the last two terms in \eqref{eq:strain_E_thermo_FEM}, denoted by $\left . (\bullet) \right |_c$, have to be kept constant in the above derivations. In this way, the weak forms in \eqref{eq:wf_mech} and \eqref{eq:wf_therm} are recovered.
 
The corresponding linearizations can be derived from \eqref{eq:Res_u_T}, yielding the tangent stiffness matrix for the coupled thermo-mechanical problem
 \beq
  \mathbf {K}_e = \begin{bmatrix} \mathbf {K}_e^{u\,u} & \mathbf {K}_e^{u\,\theta} \\
  \mathbf {K}_e^{\theta\, u} & \mathbf {K}_e^{\theta\,\theta} \end{bmatrix}
   \label{eq:K_e}
 \eeq
 with
 \beq
  \mathbf {K}^{u\,u}_e = \frac{\partial \mathbf {R}^u_e( \boldsymbol {u}_h\,,\theta_h) }{\partial \mathbf u_e} \,\,, 
   \mathbf {K}^{u\,\theta}_e = \frac{\partial \mathbf {R}^u_e( \boldsymbol {u}_h\,,\theta_h) }{\partial\boldsymbol{\theta}_e} \,\,, 
     \mathbf {K}^{\theta\,u}_e = \frac{\partial  \mathbf {R}^\theta_e( \boldsymbol {u}_h\,,\theta_h) }{\partial \mathbf u_e} \,\,, 
  \mathbf {K}^{\theta\,\theta}_e = \frac{\partial  \mathbf {R}^\theta_e( \boldsymbol {u}_h\,,\theta_h) }{\partial  \boldsymbol{\theta}_e} .
    \label{eq:K_u_T}
 \eeq
 All derivations were performed using AceGen, see \cite{KoWr16}.
 
  \subsection{Finite elements for the third medium}
The third medium acts purely in a passive manner; consequently, it is typically not subjected to external loading, and all corresponding loading terms in \eqref{eq:Potential_solid_FEM} vanish, leading for the third medium element formulations with quadratic  ansatz functions (Q2 and H2)  to
 \beq
\Pi_m( \boldsymbol {u}_h\,,\theta_h) =  \int\limits_{\Omega} \left [ \Psi_m(\boldsymbol C_h\,,\theta_h)\,\right ] \,\mbox{d}\Omega + \int\limits_{\Omega} \left [ \Psi_{R\,n}(\boldsymbol F_h)\,\right ] \,\mbox{d}\Omega
\label{eq:Potential_TM_FEM}
\eeq
where the index $n$ relates to the two-and three-dimensional case with $n=2\,,3$ and the regularization terms \eqref{eq:W_regular_all_red3} and \eqref{eq:W_reg_Hu_skew}, respectively. 

In the same way, the approximated form of the regularization \eqref{eq:reg_TMC_3d} which is employed for two- and three-dimensional linear Q1, H1  and T1  elements yields 
 \beq
\Pi_m( \boldsymbol {u}_h\,,\theta_h) =  \int\limits_{\Omega} \left [ \Psi_m(\boldsymbol C_h\,,\theta_h)\,\right ] \,\mbox{d}\Omega + \int\limits_{\Omega} \left [ \Psi_{Rf}(\boldsymbol u_h\,,p_{h\,i})\,\right ] \,\mbox{d}\Omega
\label{eq:Potential_TM_FEM_appr}
\eeq
As discussed in the preceding sections, the strain energy in the third medium must remain sufficiently small to prevent any unintended influence on the actual deformation behavior.
The coupled strain energy \eqref{eq:strain_E_thermo_FEM} remains the same, however the conductivity has to be adjusted for the thermomechanical contact. The idea is to define a small conductivity $k_{TM}$ for the third medium  to model the conductivity for a gas (air) around the contacting solids.  In the light of these physical models we define a conductivity of the third medium as provided in \eqref{eq:Fourier_law_TMC}.

The displacement field $\boldsymbol u$ and the temperature field  $\theta$ of the third medium  are approximated  in the same way as the related fields for the solids, see \eqref{eq:discr_u_theta}. 
The potentials \eqref{eq:Potential_TM_FEM} and \eqref{eq:Potential_TM_FEM_appr} are formulated without a loading term, see above. However, there is  one  exception which is related to  pressure or suction loading of cavities that can be directly applied to the third medium, see  \cite{FaHoDoRo24} and \cite{WrKoJu25}.

The gradients of the displacement and temperature  field are discretized as in \eqref{eq:F_h} and \eqref{eq:grad_q_h}.  The field variables $p_i$ for the regularization in \eqref{eq:reg_TMC_3d_beta} are approximated  by the same ansatz functions $N_I$ that are employed to approximate displacements and temperature  
   \beq
 {p}_{i\,h} = \sum_{K=1}^{n}\,N_K(\boldsymbol{\xi}) \,{p}_{i\,K }\,.
 \label{eq:pi_h}
 \eeq
Since all fields are continuous the element contributions can be assembled in a standard way.

 The finite element code for each of the elements is automatically generated by employing the  tool AceGen. Performing again the same derivations as in \eqref{eq:Res_u_T} and \eqref{eq:K_u_T} yield the residual vector and the tangent matrix of the third medium element.


 \section{Numerical examples}
 \label{sec:Numerics}
With two- and three-dimensional  examples, we illustrate the accuracy and efficiency of the new formulations of the third medium using linear shape function. 
 The element types, used in this paper are 
 \begin{itemize}
 \item triangular elements (2d) with linear ansatz functions (T1) and three nodes,
\item quadrilateral elements (2d) with linear ansatz functions (Q1) and six nodes,
   \item hexahedral elements (3d) with linear ansatz functions (H1) and eight nodes and
      \item hexahedral elements (3d) with quadratic  ansatz functions (H2 and H2S) and 27 and 20 nodes.
 \end{itemize}
 All computations were performed with the software tool AceFEM, see \cite{Kor00b} and \cite{KoWr16}, on a  notebook with M4 processor.

 \subsection{Two-dimensional examples}  
 
  Three numerical examples will be discussed in order to show the performance of linear triangular (T1) and quadrilateral (Q1) elements as a third medium.  All stabilization parameters are selected as discussed in Section \ref{sec:reduced_grad}.
 
 \subsubsection{Block under vertical load with heat conduction}
 
\begin{figure}[h!]
	\centering
	\includegraphics[width=0.51\textwidth]{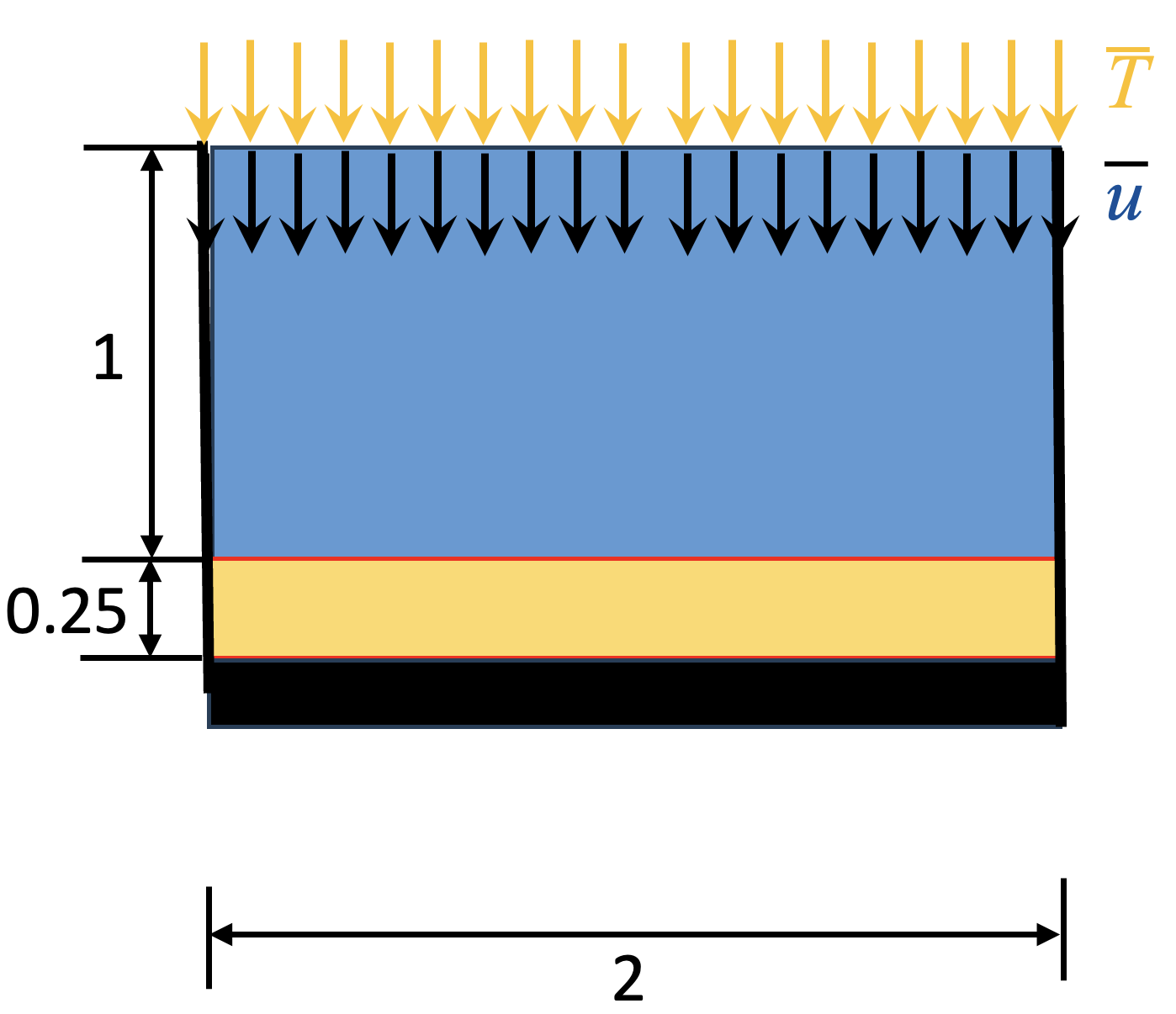}
	\caption{Block under displacement and thermal load}
	\label{fig:single_block}
\end{figure}

 An example which has an analytical solution is discussed first to validate the thermo-mechanical contact formulation. The system is provided in Fig. \ref{fig:single_block} together with the loading. The upper part of dimensions $1 \times 2$ is the solid while the lower part with dimensions  $0.25 \times 2$ is the third medium.
The block is fixed in $x$- and $y$-direction at the bottom and on the top in $x$-direction. 
The material parameters for the structural components are the modulus of compression $K=20$ and the shear modulus $\mu=10$. The thermal conduction of the solid is given by $k_\theta= 100$ while the conductance of the third medium is $k_{TM}=1$. The value for the stiffness of the third medium was selected as $\gamma=10^{-4}$ and the regularization parameter is given by $\beta_2=10^{-2}$. 

The applied load is a prescribed uniformly distributed vertical displacement of $\bar u = -0.4$. The temperature at the top is given by $\bar T = 100$ and at the bottom by $\bar T =20$.

\begin{figure}[h!]
	\centering
	\includegraphics[width=0.42\textwidth]{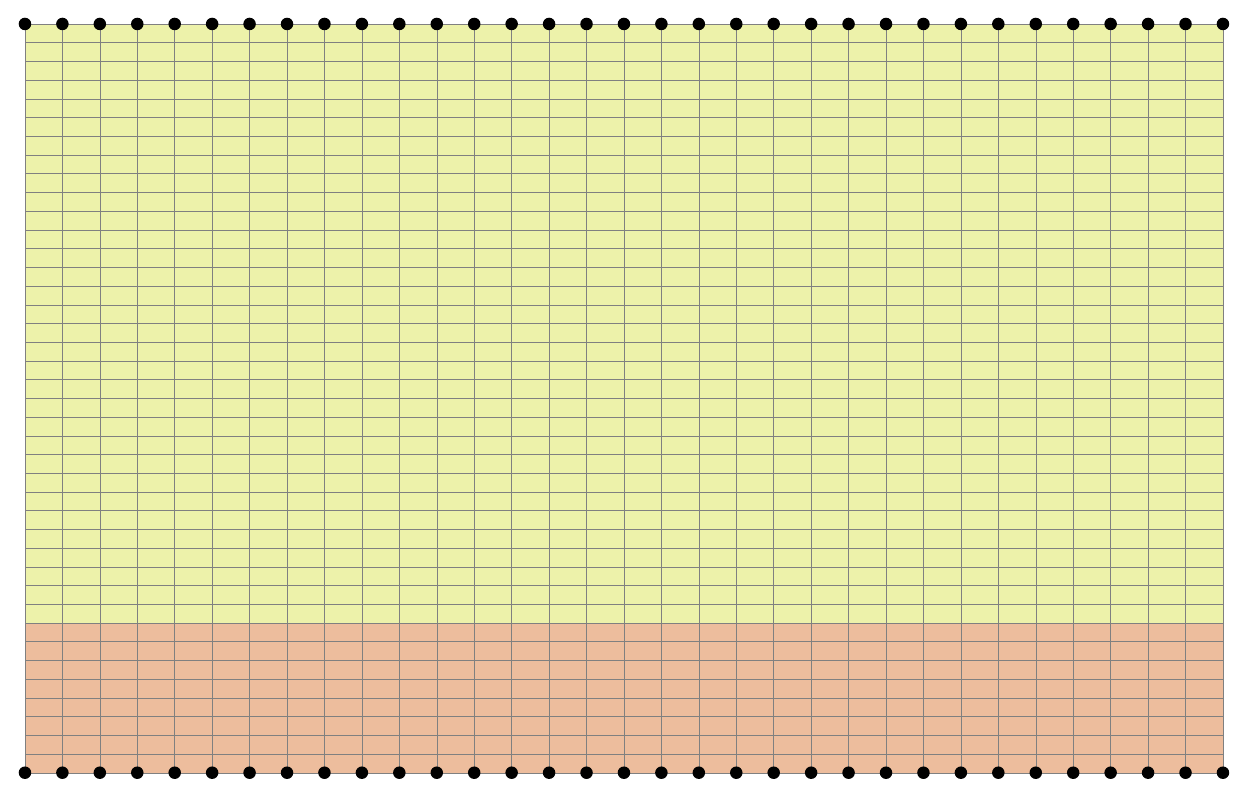}
	\includegraphics[width=0.50\textwidth]{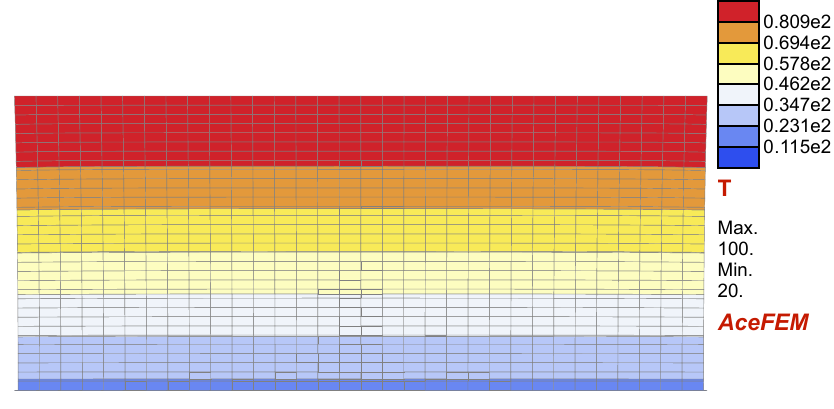}
	\caption{Block: mesh and deformed solid with temperature distribution.}
	\label{fig:single_block_mesh + defo}
\end{figure}

Fig. \ref{fig:single_block_mesh + defo} depicts the mesh consisting of quadrilateral Q1-elements with density  $32 \times 32$ for the solid and $32 \times 8$ for the third medium together with the final deformed state that shows the temperature distribution 
Observe that the third medium is totally sqeezed together. Note, that there is no bulging of elements in the third medium. Thus, the regularization works perfectly in this case. 

\begin{figure}[h!]
	\centering
	\includegraphics[width=0.5\textwidth]{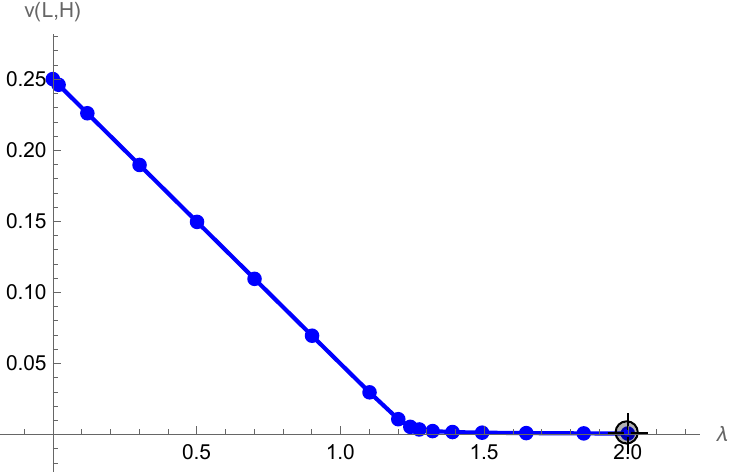}
	\caption{Block: Load-gap curve.}
	\label{fig:Load_gap}
\end{figure}

The gap $g$ is plotted versus the loading parameter  in Fig. \ref{fig:Load_gap}. The final gap is
$g=6.3 \cdot10^{-4}$. 

\begin{figure}[h!]
	\centering
	\includegraphics[width=0.48\textwidth]{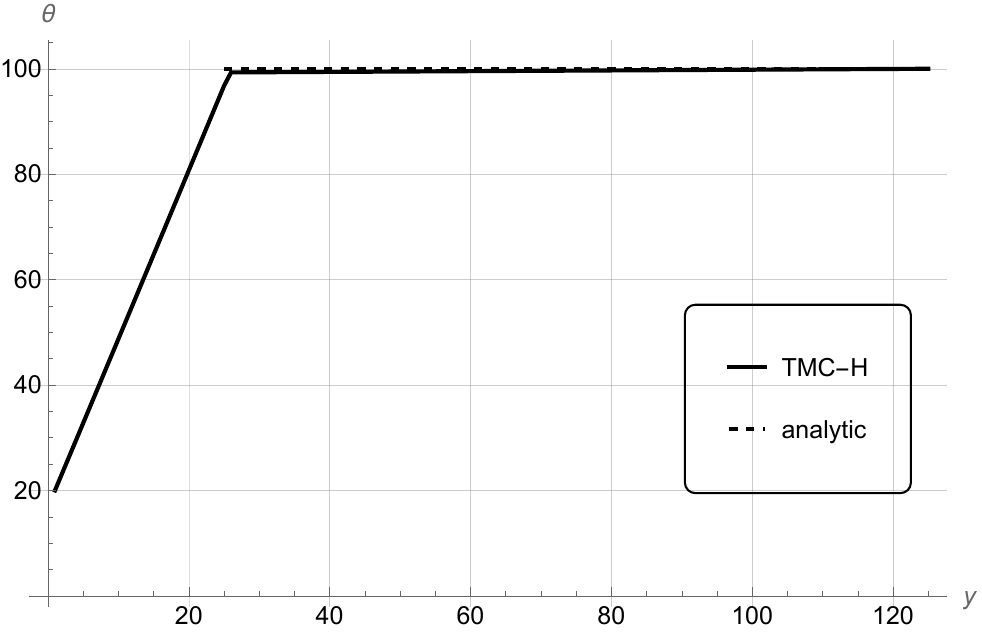}
	\includegraphics[width=0.48\textwidth]{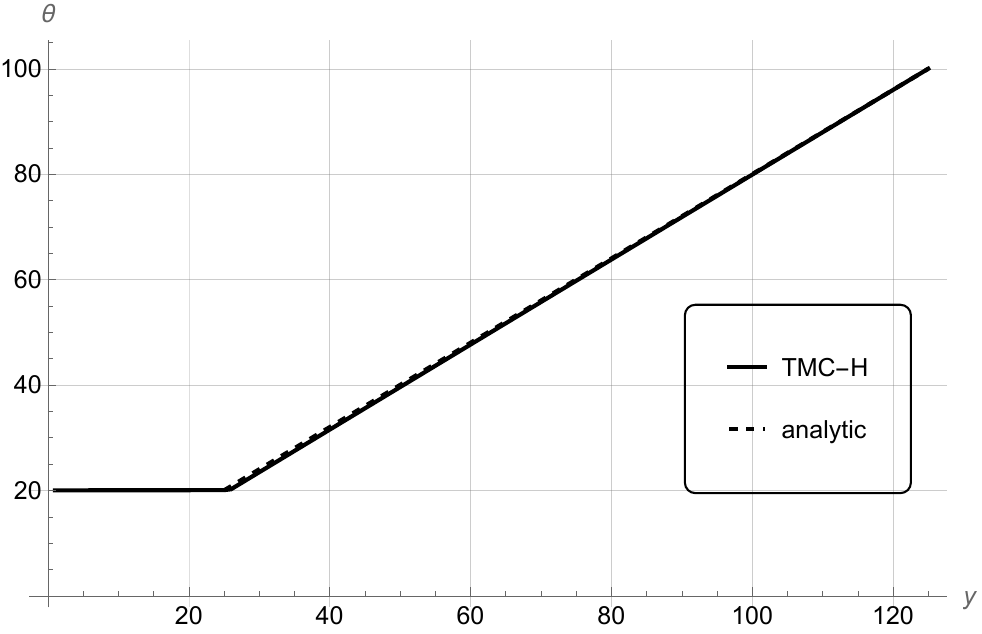}
		\caption{Block: progression of the temperature just before and after contact.}
	\label{fig:Temperature_progression}
\end{figure}

The temperatur distribution has two stages. Until the gap is closed, the temperature has a linear progression in the third medium while maintaining the applied temperature of $\bar T = 100$ in the solid, see left side in Fig. \ref{fig:Temperature_progression}. Then after contact the temperature distribution has a  linear progression in the solid from $T=20$ to $T=100$ as expected, see right side in Fig.\ref{fig:Temperature_progression}. The results match the analytical solution.

 \subsubsection{Two blocks under partial vertical load with heat conduction}

\begin{figure}[h!]
	\centering
	\includegraphics[width=0.4\textwidth]{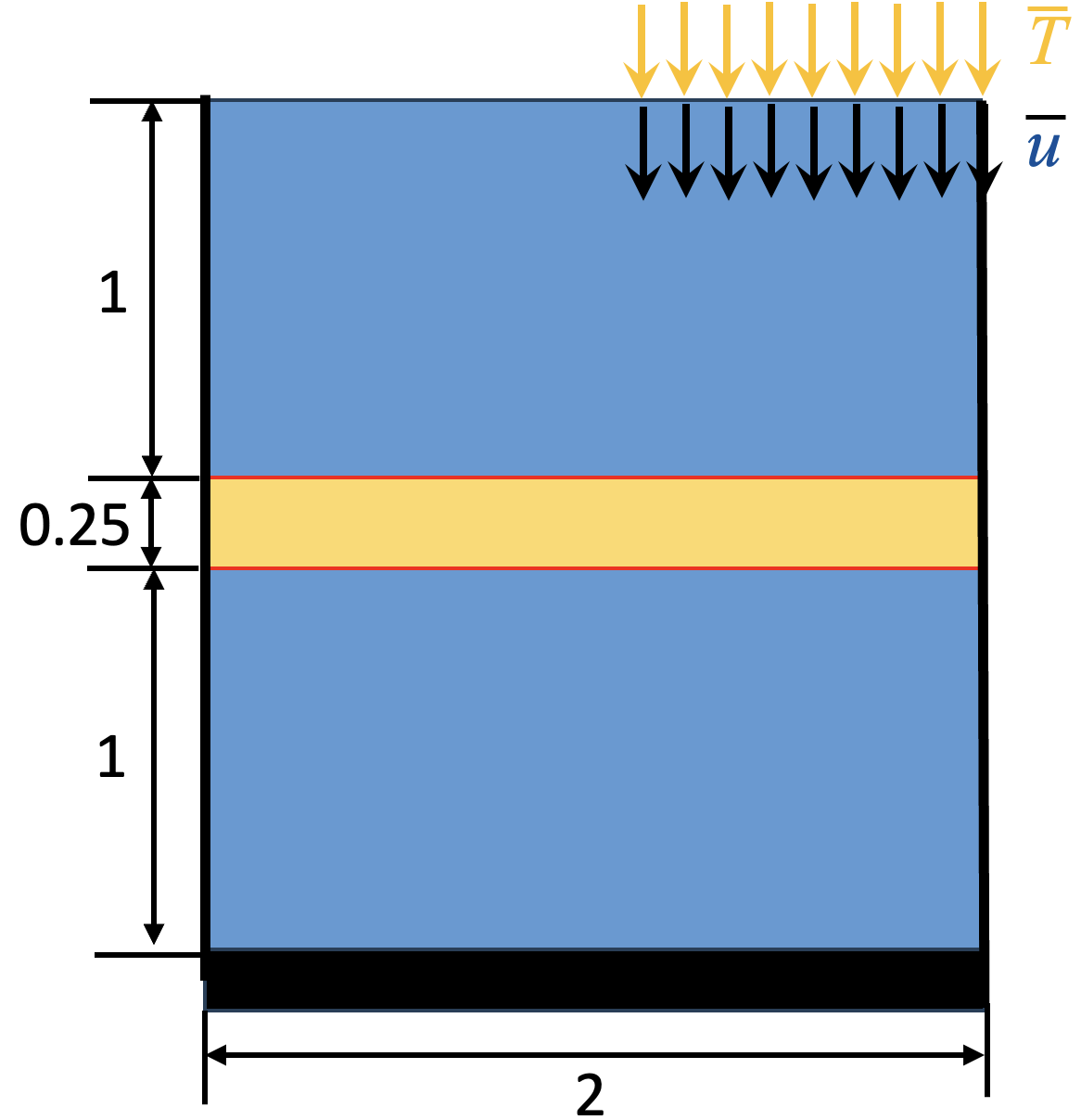}
	\includegraphics[width=0.32\textwidth]{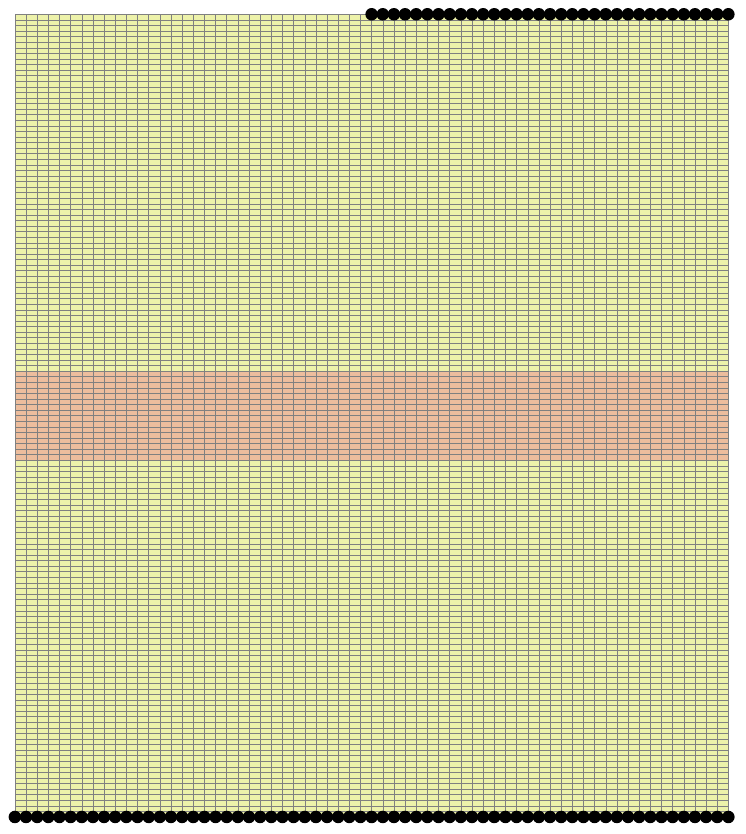}
	\caption{Two blocks seperated by the third medium and finite element mesh.}
	\label{fig:2_blocks_system}
\end{figure}

The two blocks are separated by the third medium as depicted on the left side in Fig. \ref{fig:2_blocks_system} while the right side shows the used discretization with quadrilateral  Q1-elements. The lower block is fixed in $x$- and $y$-direction at the bottom and the upper block is fixed on the right half in $x$-direction.  The temperature at the bottom of the lower block is assumed to be zero.
The material parameters for the structural components are the modulus of compression $K=20$ and the shear modulus $\mu=10$. The thermal conduction of the solid is given by $k_\theta= 10$ while the conductance of the third medium is $k_{TM}=0.1$. The value for the stiffness of the third medium was selected as $\gamma=10^{-4}$ and the regularization parameter is given by $\beta_2=10^{-2}$. 
Both solids are discretized by $64 \times 64$ triangular elements and the third medium by $64 \times 16$ elements which yields a mesh with a total of 9216 elements, 18850 nodes and 37406 degrees of freedom.

\begin{figure}[h!]
	\centering
	\includegraphics[width=0.4\textwidth]{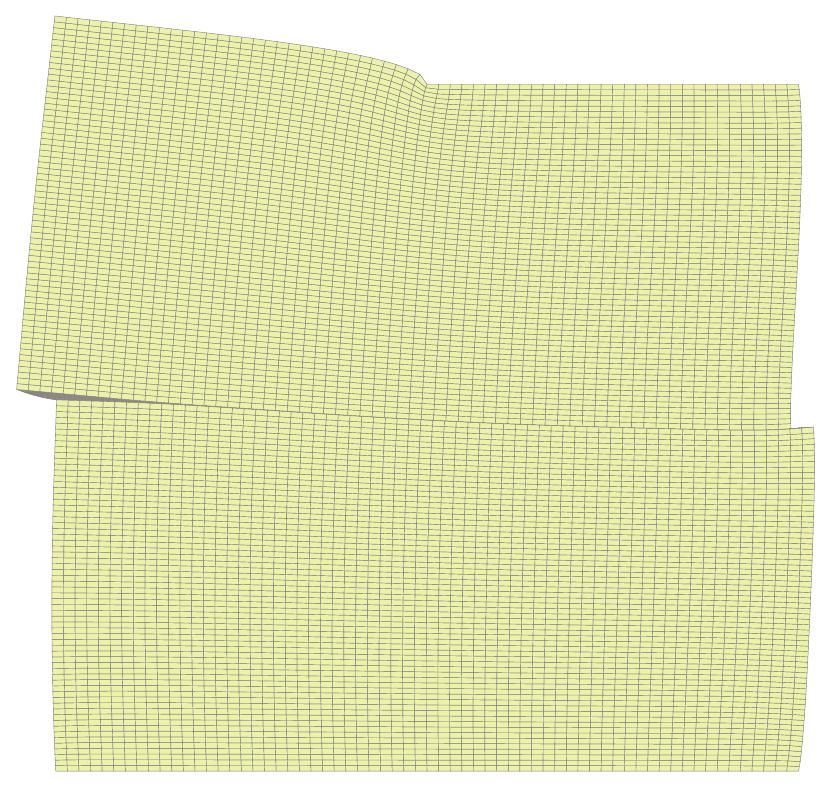}
	\includegraphics[width=0.45\textwidth]{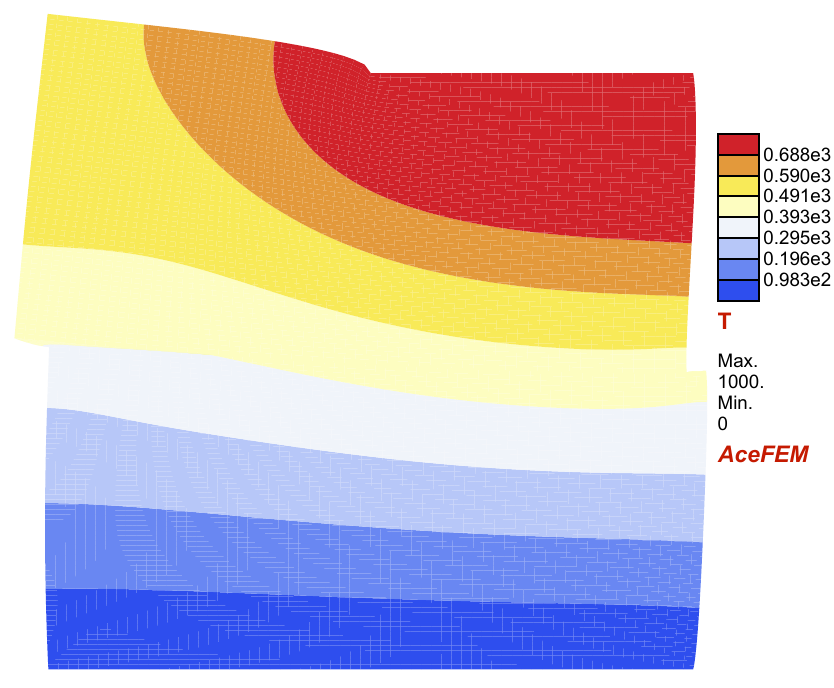}
	\caption{Deformed configuration and temperature distribution of two blocks under vertical loading.}
	\label{fig:2_blocks_defo+T}
\end{figure}

The deformed configuration is computed for a prescribed vertical displacement of  $\bar u = 0.4$ together with temperature of $\bar T= 1000$ at the top right half of the upper block, see left side of  Fig. \ref{fig:2_blocks_system}. Figure   \ref{fig:2_blocks_defo+T} shows the deformed mesh on the left side and the temperature distribution on the deformed mesh on the right side without the mesh of the third medium. 
It can be seen that the third medium is extremely compressed but also sheared due to the tangential deformation. However, there is no bulging of the third medium visible. The solution needs only 15 steps with a total of 93 iterations. Also the temperature distribution related to the compressed state is well recovered with a basically linear distribution in the deformed state. 

\begin{figure}[h!]
	\centering
	\includegraphics[width=0.45\textwidth]{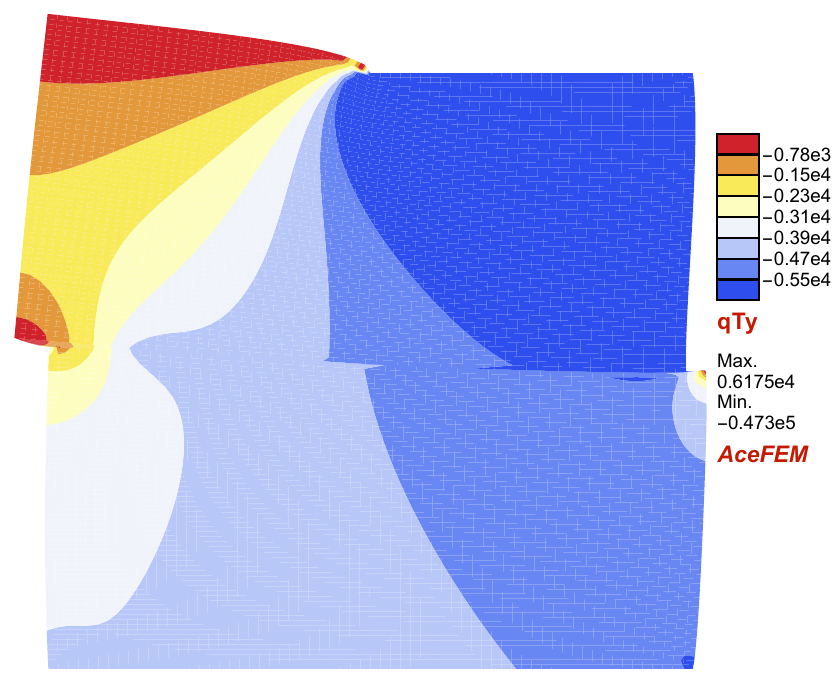}
	\includegraphics[width=0.45\textwidth]{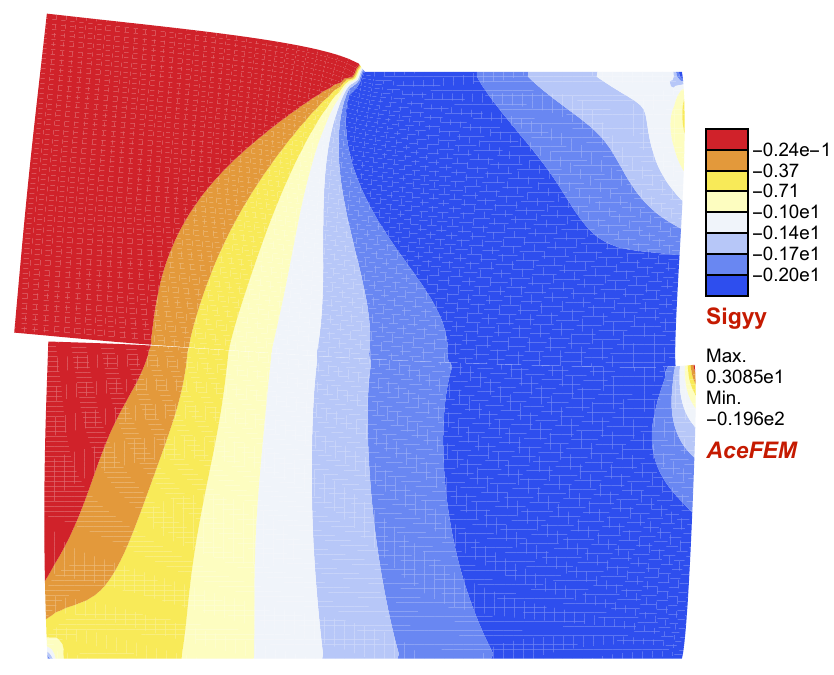}
	\caption{Heat transfer component $q_y$ (left) and Cauchy stress $\sigma_{yy}$ (right).}
	\label{fig:2_blocks_defo+qTy+Syy}
\end{figure}

The distribution of the vertical component  of the heat flux, $q_y$, and the component of the Cauchy stresses $\sigma_{yy}$  is illustrated in Fig. \ref{fig:2_blocks_defo+qTy+Syy}.
For better visibility of the contact area the deformed mesh of the third medium is not plotted.
Note that the strong tangential deformation at the contact interface does not affect the smooth transition of the stress field across the contact zone. This behaviour can also be observed for the heat transfer.

 \subsubsection{Contact between two discs and an elliptical body}
 \label{sec:two_discs}

\begin{figure}[h!]%
	\centering
	\includegraphics[width=0.49\textwidth]{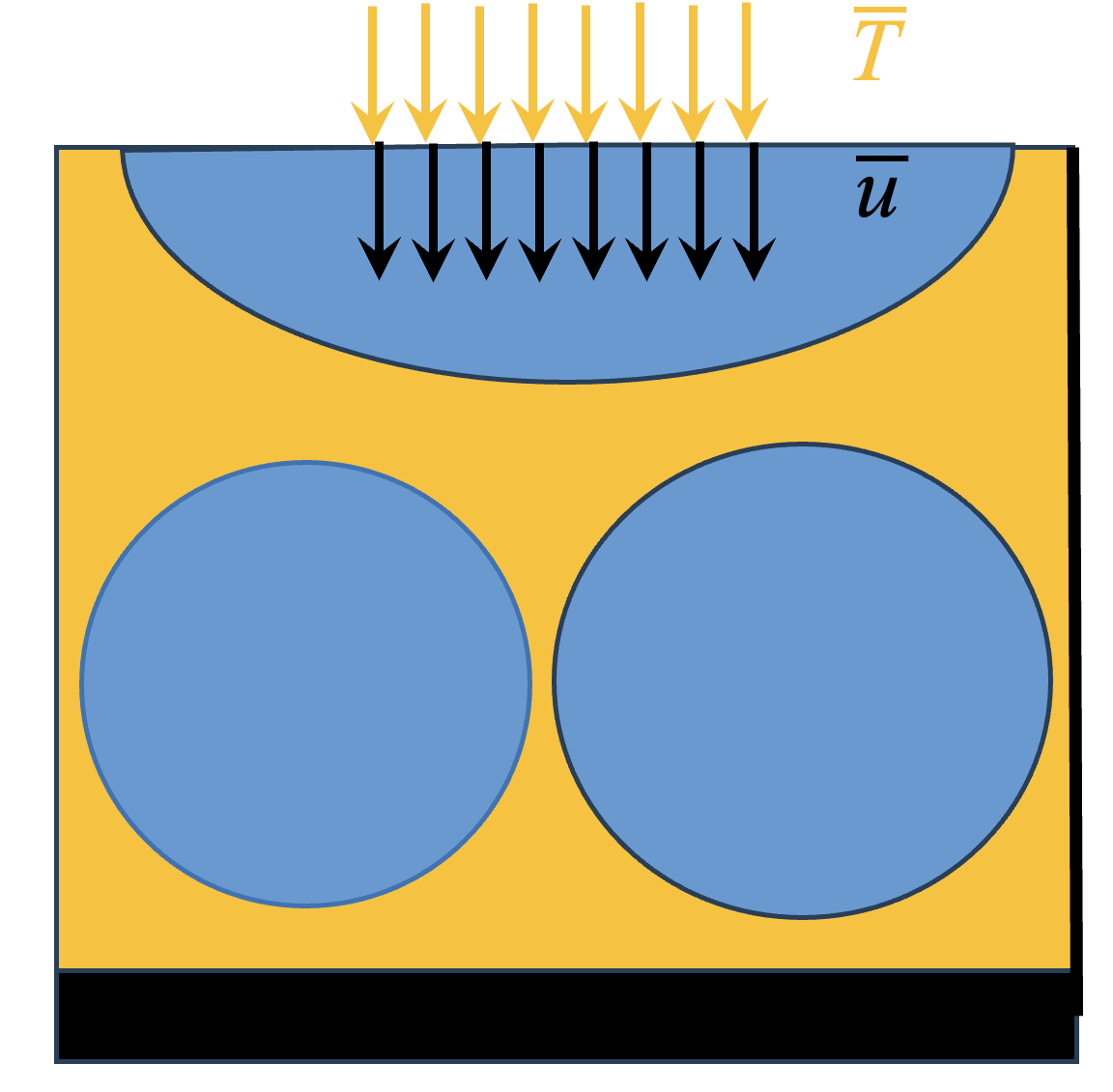}
	\includegraphics[width=0.49\textwidth]{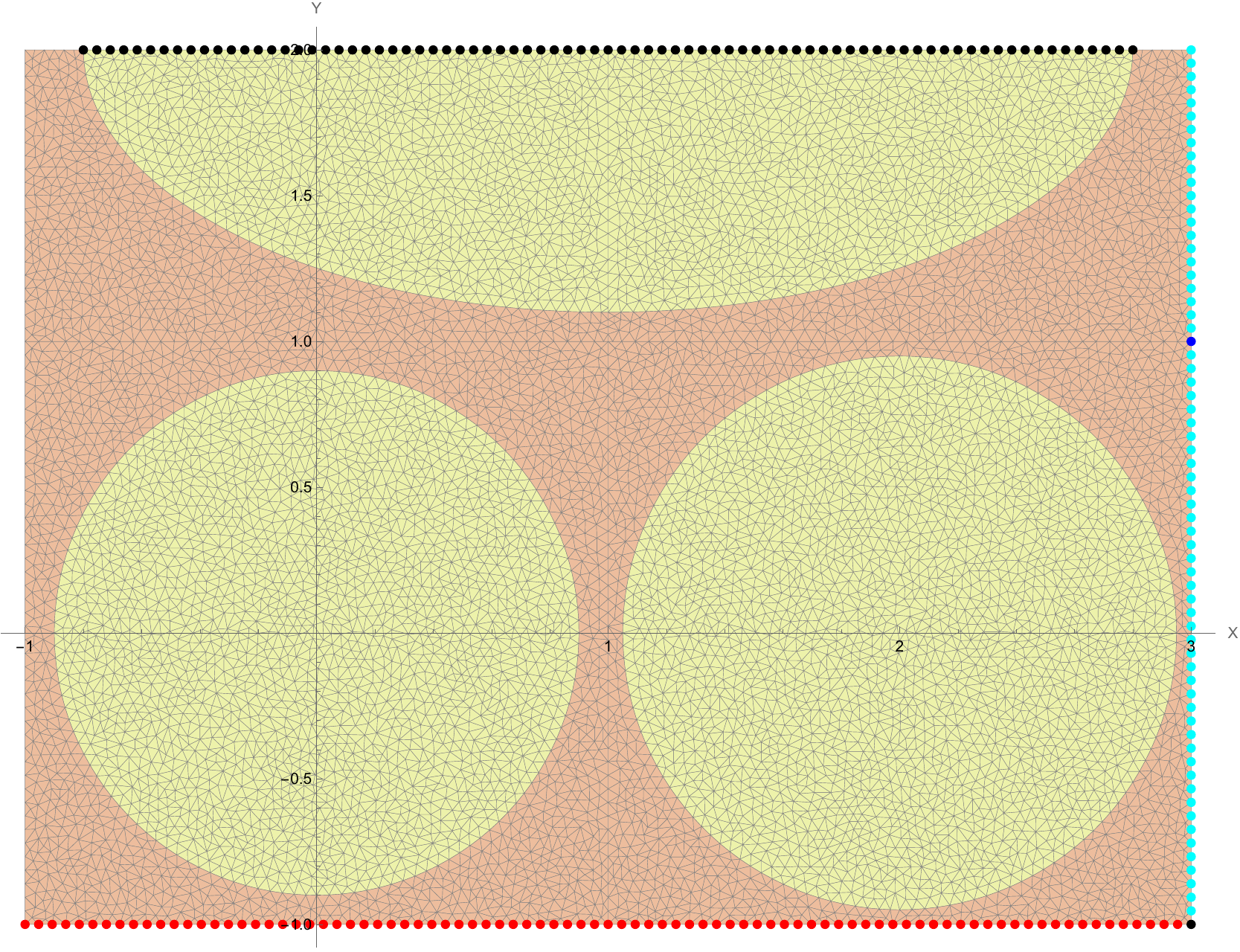}
	\caption{System of the two disks and an ellipse embedded in  the third medium, left dimension and right the  mesh with coordinate system.}
	\label{fig:discs_ellipse_system}
\end{figure}

Three solids of different size are surrounded by the third medium which fills out a box as shown in the left part of Fig. \ref{fig:discs_ellipse_system}.  The box has a width of $B=4$ snd a height of $H=3$. The associated finite element mesh consists of 18643 linear triangular T1-elements and is depicted on the right side of Fig. \ref{fig:discs_ellipse_system} with a total number of 37430 unknowns..
The radii of the left and right disk are $r_{left}= 0.9$ and  $r_{right}= 0.95$, respectively. The boundary of the half ellipse is defined by the function $(x-1)^2/4 +(y-2)^2 =0.9^2$. The modulus of compression $K$ and the shear modulus $\mu$ are equal for all three solids. The values are $K=20$ and $\mu=10$. The conductivities of the solids and the third medium is  $k_\theta=1000$  and $k_{TM}=1$, respectively. The thermal expansion coefficient for the solids and the third medium is $\alpha_t= 10^{-4}$ and the stiffness of the third medium is $\gamma= 10^{-4}$ while the constant $\beta_2$ has the value $10^{-2}$.

\begin{figure}[h!]%
	\centering
	\includegraphics[width=0.95\textwidth]{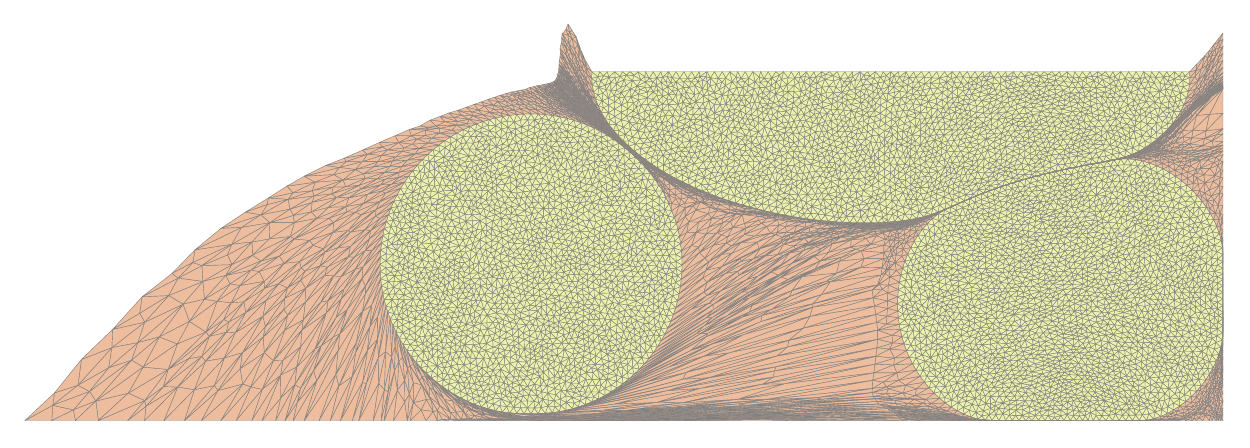}
	\caption{Deformed discs and ellipse.}
	\label{fig:Sp_ellipse_defo}
\end{figure}

\begin{figure}[h!]%
	\centering
	\includegraphics[width=0.49\textwidth]{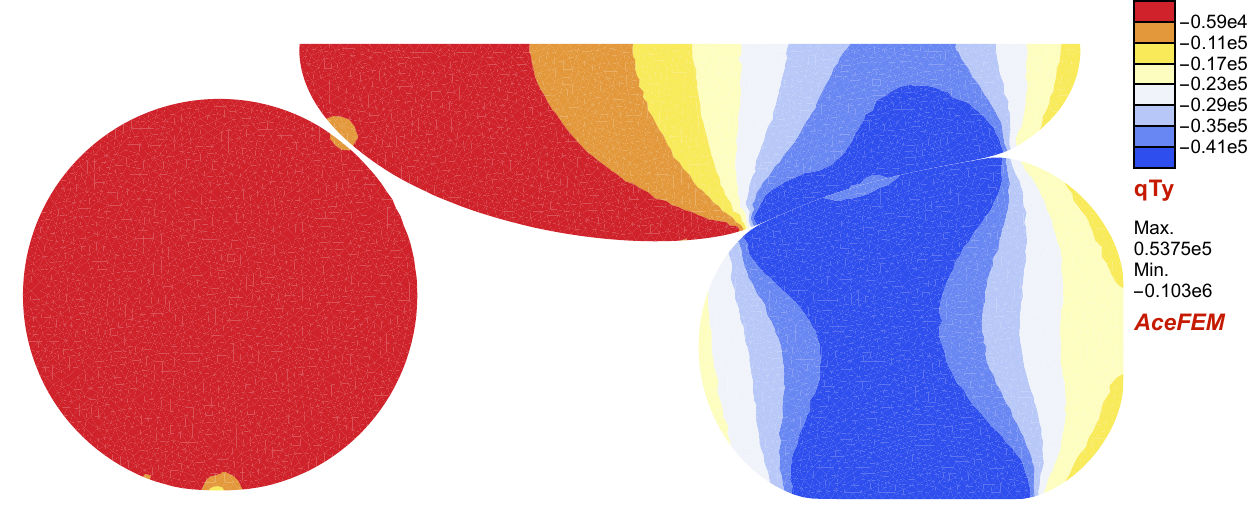}
	\includegraphics[width=0.49\textwidth]{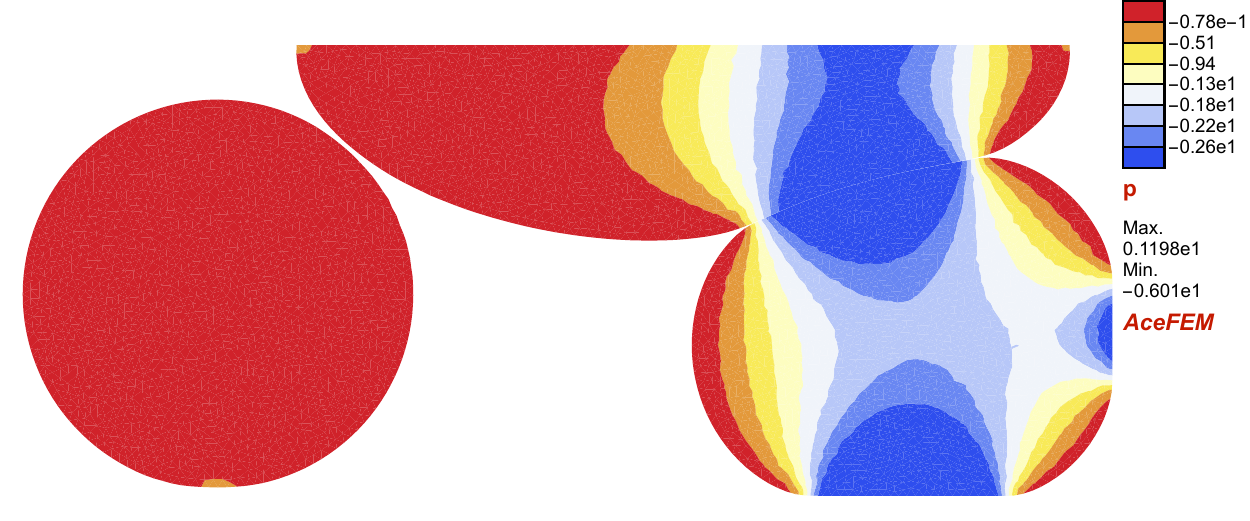}
	\caption{Heat conduction in vertical direction (left side) and contact pressure (reight side), plotted without showing the deformed mesh of the third medium.}
	\label{fig:Sp_ellipse_qTy_p+defo}
\end{figure}

A temperature of $\bar T=90$ is applied together with a vertical displacement of $\bar u = 0.9$ at the top of the ellipse. The automatic load adaptive procedure leads to a total of 103 steps with 556 iterations in order to obtain the deformed configuration in Fig. \ref{fig:Sp_ellipse_defo}.
Since there is no boundary constraint on the right side, the left sphere is displaced outward - a consequence of the absence of friction in the model. As a result, significant deformations occur in the underlying third medium on the left side. However, these deformations do not influence the thermal or mechanical fields. This is evident in Fig. \ref{fig:Sp_ellipse_qTy_p+defo}, which shows the vertical heat flux $q_y$ and the contact pressure $p$. For clarity, the third medium as well as the mesh  is omitted from the figure, allowing a clearer view of the solid deformations, contact interactions, and the distributions of $q_y$ and $p$. 
It is worth noting that a very small amount of heat is transferred to the left free disc, primarily due to thermal conduction through the third medium, which possesses a low thermal conductivity $k_{TM}$, see \eqref{eq:Fourier_law_TMC}. 
From a mechanical perspective, however, the third medium does not transmit any contact pressure to the left disc. Despite the non-matching mesh at the contact interfaces with the highly compressed third medium mesh, both the contact pressure distribution and the heat flux remain continuous across the interface as shown in Fig. \ref{fig:Sp_ellipse_qTy_p+defo}.

 \subsubsection{Heat conductance at micro-scale}

\begin{figure}[h!]%
	\centering
	\includegraphics[width=0.4\textwidth]{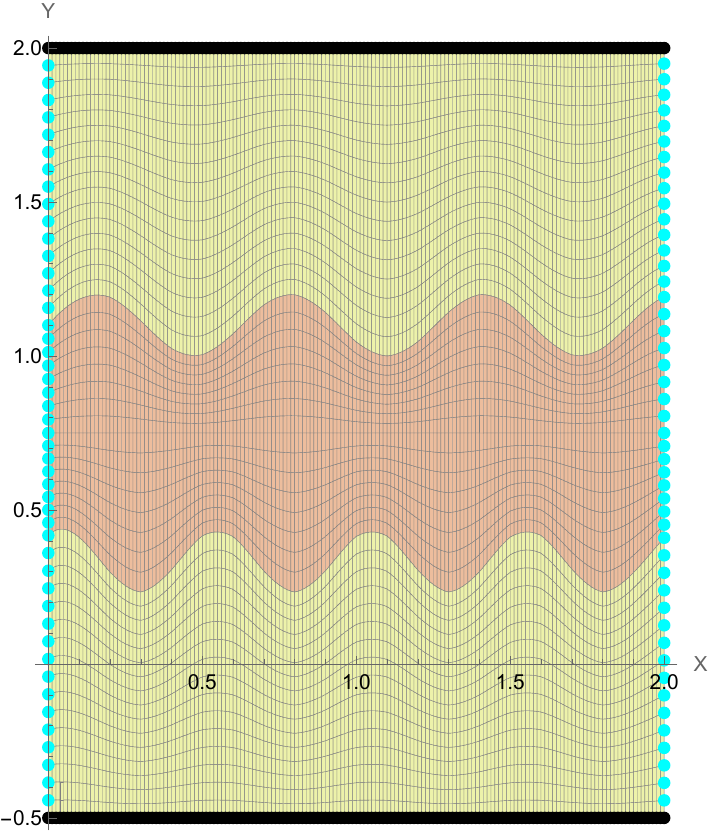}
	\caption{Micro-scale system with wavy surfaces at the contact interface.}
	\label{fig:TMC_micro_system}
\end{figure}

Contact conductance is particularly important when surfaces are in physical proximity but separated by micro-gaps or third media, where real contact occurs only at discrete asperities. In such cases, thermal conduction is governed by the actual contact area, the properties of the interstitial medium (e.g., gas, vacuum, or a soft solid), and the surface topography. The following example, in which two soft solid surfaces are connected via the third medium, illustrates various contact heat transfer mechanisms, including gas-phase conduction and localized spot conductance. The mesh is depicted in Fig. \ref{fig:TMC_micro_system} together with the axis-system and the boundary conditions.
At the bottom all displacements are fixed in $x$- and $y$-direction and the temperature is set to zero. The displacement components in $x$-direction are fixed at both sides of the system. At the top, the all displacements are fixed in $x$-direction. A temperature of $\bar T=100$ is prescribed at the top surface as well as a vertical diplacement of maximum $\bar v=4/3$.

\begin{figure}[h!]%
	\centering
	\includegraphics[width=0.32\textwidth]{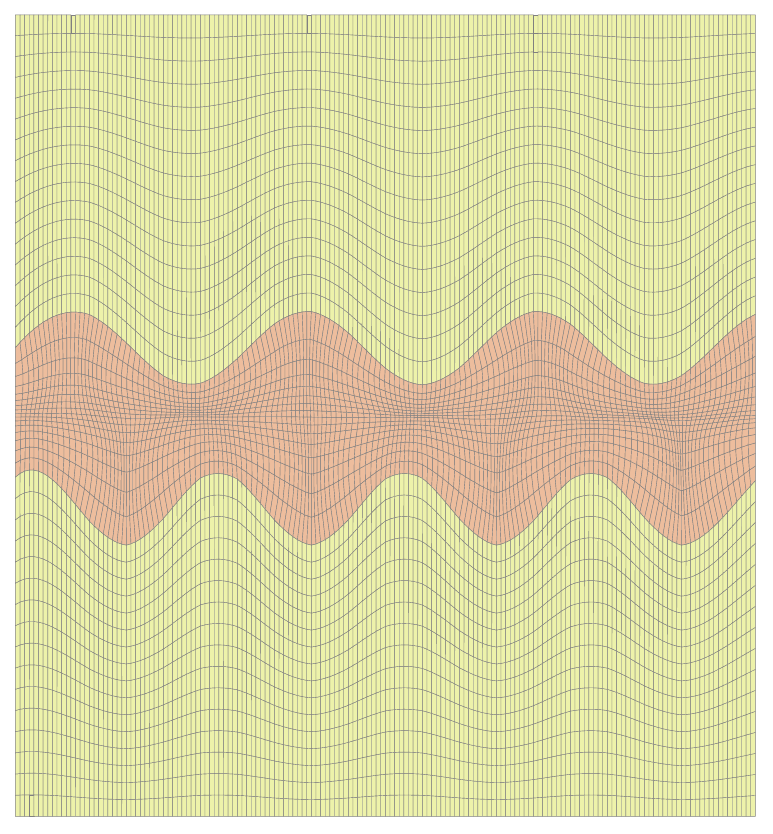}
	\includegraphics[width=0.32\textwidth]{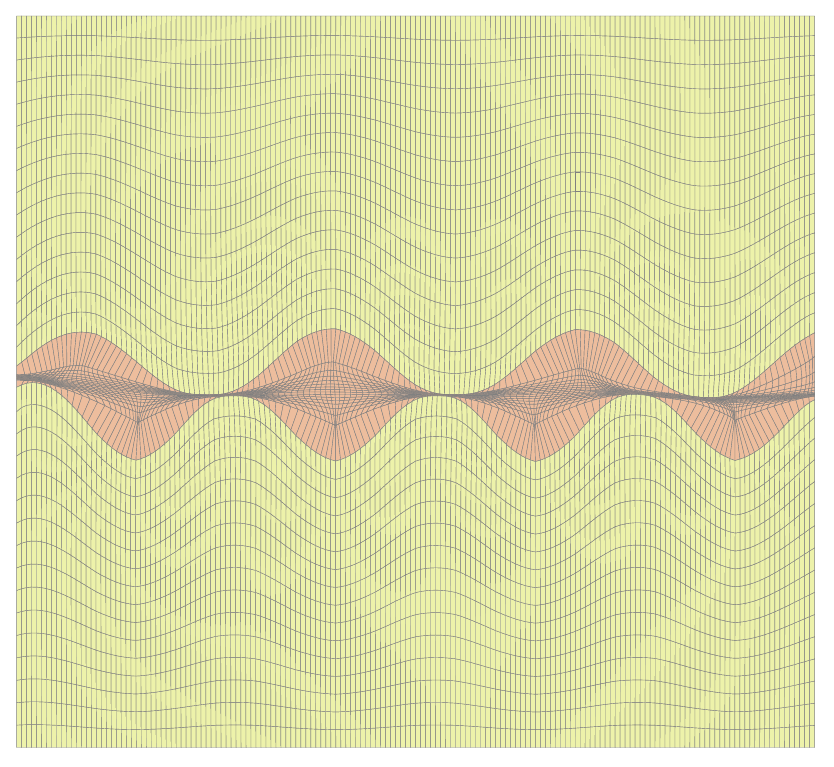}
	\includegraphics[width=0.32\textwidth]{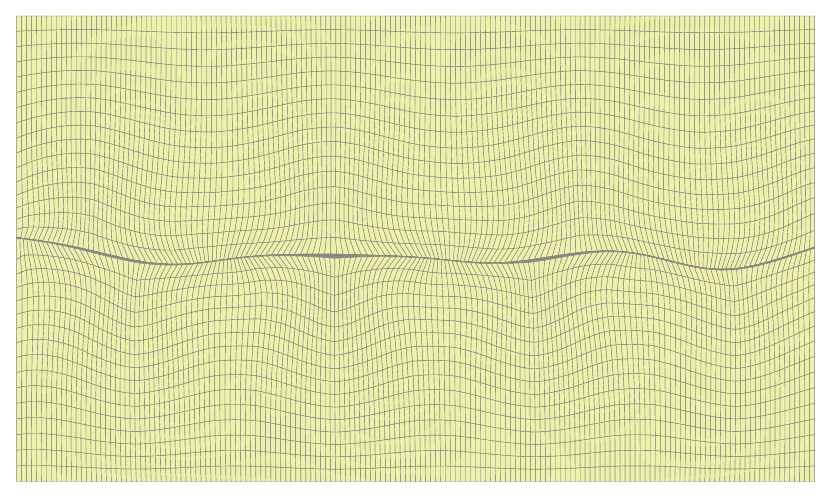}
	\caption{Deformed configuration for $\bar v = 1/3$, $\bar v = 2/3$ and $\bar v = 4/3$.}
	\label{fig:TMC_micro_defos}
\end{figure}

The deformed meshes are depicted in Fig. \ref{fig:TMC_micro_defos} for three different prescribed displacements.
Clearly for the prescribed displacement of $\bar v = 1/3$ there is no contact, while for $\bar v = 2/3$ contact occurs in the two left spots. Finally $\bar v = 4/3$ defines a full contact state with a completely compressed third medium.

\begin{figure}[h!]%
	\centering
	\includegraphics[width=0.32\textwidth]{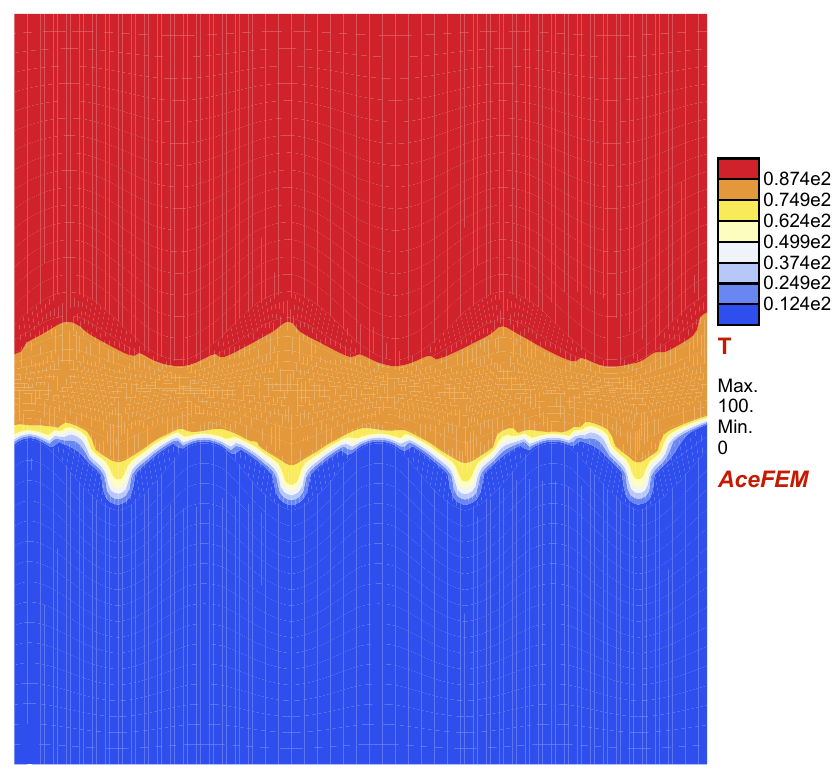}
	\includegraphics[width=0.32\textwidth]{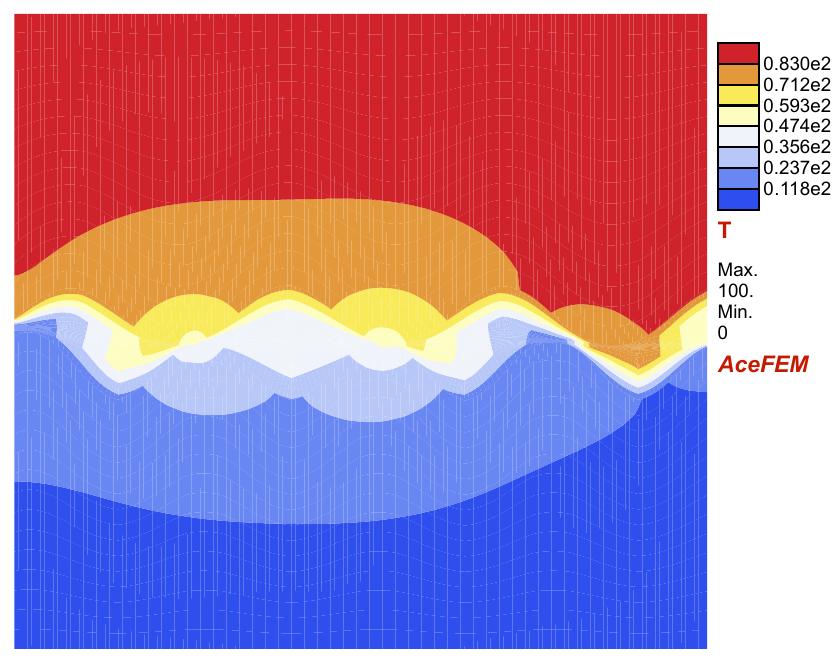}
	\includegraphics[width=0.32\textwidth]{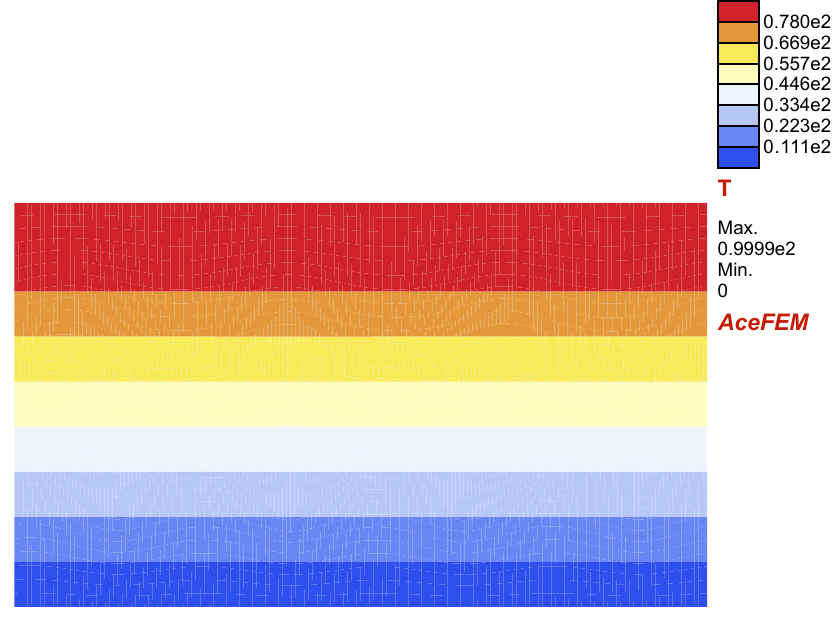}
	\caption{Deformed configuration for $\bar v = 1/3$, $\bar v = 2/3$ and $\bar v = 4/3$.}
	\label{fig:TMC_micro_T}
\end{figure}

Figure \ref{fig:TMC_micro_T} presents the temperature distributions for both configurations, clearly illustrating the distinct phases of heat conduction. To enhance visual clarity, the computational meshes are omitted, allowing an unobstructed view of the temperature field. On the left, for the prescribed vertical displacement of $\bar{v} = 1/3$, a pronounced temperature discontinuity is observed, indicating the low thermal conductance of the third medium at this stage.
At a prescribed vertical displacement of $\bar{v} = 2/3$, initial contact occurs at a few asperities, enabling localized heat conduction and resulting in a different temperature distribution. As the displacement increases further to $\bar{v} = 4/3$, full contact across the interface is established, leading to a nearly linear temperature distribution over the height of the sample.

\begin{figure}[h!]%
	\centering
	\includegraphics[width=0.32\textwidth]{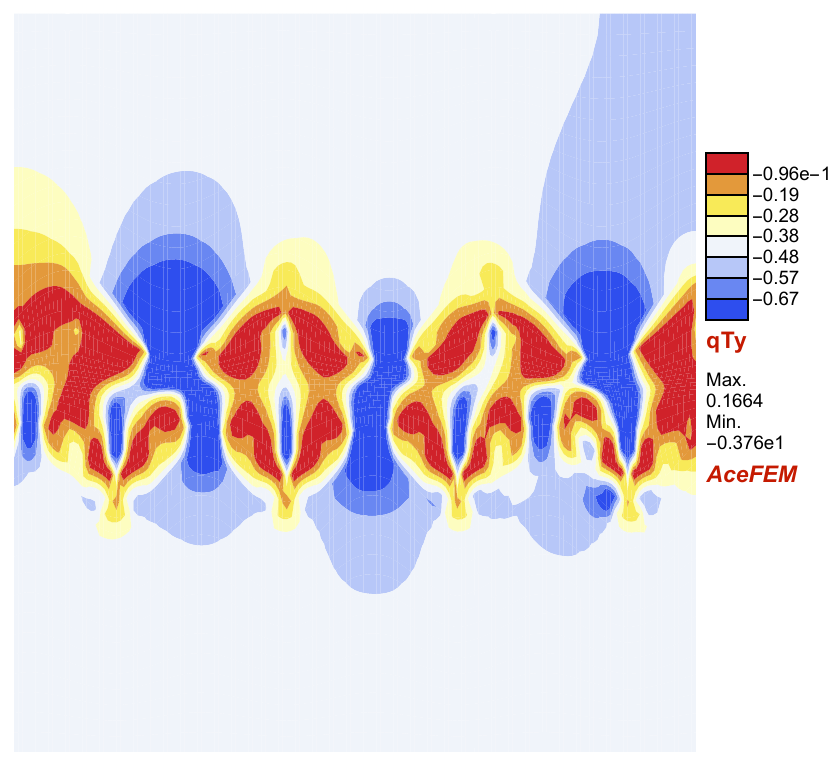}
	\includegraphics[width=0.32\textwidth]{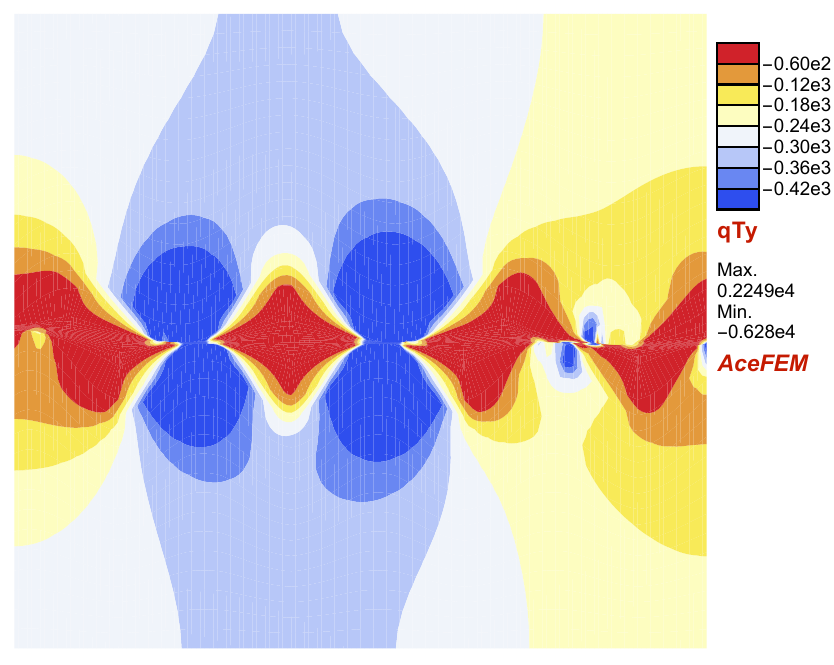}
	\includegraphics[width=0.32\textwidth]{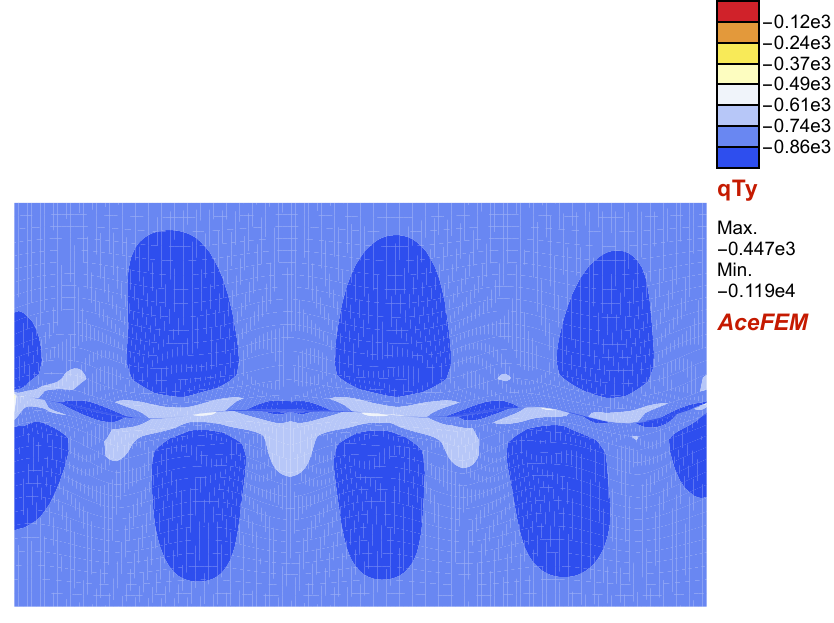}
	\caption{COmponent of the heat flux $q_y$  for $\bar v = 1/3$, $\bar v = 2/3$ and $\bar v = 4/3$.}
	\label{fig:TMC_micro_qTy}
\end{figure}

The variation in heat conductance corresponding to different contact states is also evident in Fig. \ref{fig:TMC_micro_qTy}, which displays the $q_y$ component of the heat flux. Again the meshes are omitted for clarity of the results. On the left, the interface remains open, and the heat flux through the third medium is minimal, as indicated by the low values in the legend. As initial contact occurs at isolated asperities, localized regions of elevated heat flux become clearly visible. In the final stage, when full contact is established, the heat flux becomes nearly uniform across the interface, as shown in the right portion of Fig. \ref{fig:TMC_micro_qTy}.

 \subsection{Three-dimensional example} 
This section discusses the results obtained using a hexahedral three-dimensional third medium element. The formulations employed are as discussed in the previous sections and are based on linear ansatz functions. The hexadedral element (H1) has 8 nodes. Gauss integration with a $2 \times 2 \times 2$ rule is employed.

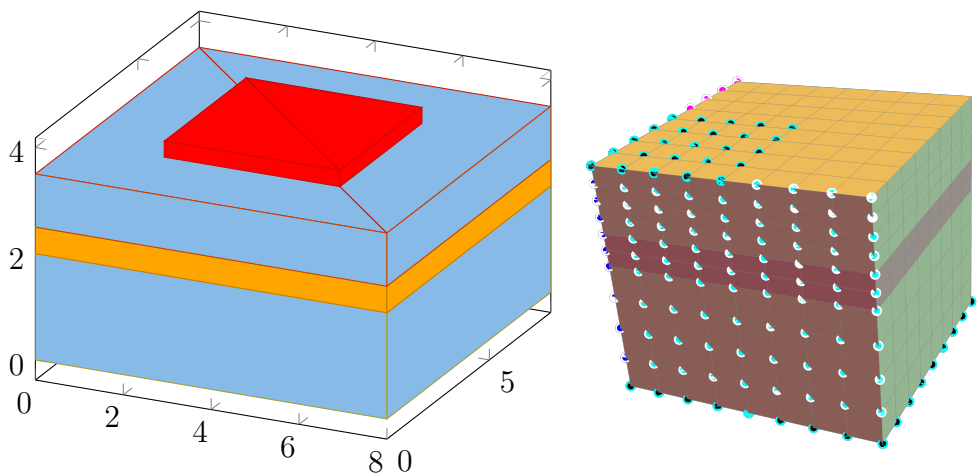
\begin{figure}[h!]%
     \begin{center}
 \begin{tikzpicture}
\begin{axis}
\addplot3 [surf,myblue] coordinates {
(0,0,0) (8,0,0) (8,8,0) 

(0,0,2) (8,0,2) (8,8,2)

(0,0,2)(0,8,2)(8,8,2)

};
\addplot3 [surf,myorange] coordinates {
(0,0,2) (8,0,2) (8,8,2) 

(0,0,2.5) (8,0,2.5) (8,8,2.5)

(0,0,2.5)(0,8,2.5)(8,8,2.5)

};
\addplot3 [surf,myblue] coordinates {
(0,0,2.5) (8,0,2.5) (8,8,2.5) 

(0,0,3.5) (8,0,3.5) (8,8,3.5)

(0,0,3.5)(0,8,3.5)(8,8,3.5)

};
\addplot3 [surf,myred] coordinates {
(2,2,3.5) (6,2,3.5) (6,6,3.5) 

(2,2,3.8) (6,2,3.8) (6,6,3.8)

(2,2,3.8)(2,6,3.8)(6,6,3.8)

};
\end{axis} 
\end{tikzpicture}
\includegraphics[width=0.4\textwidth]{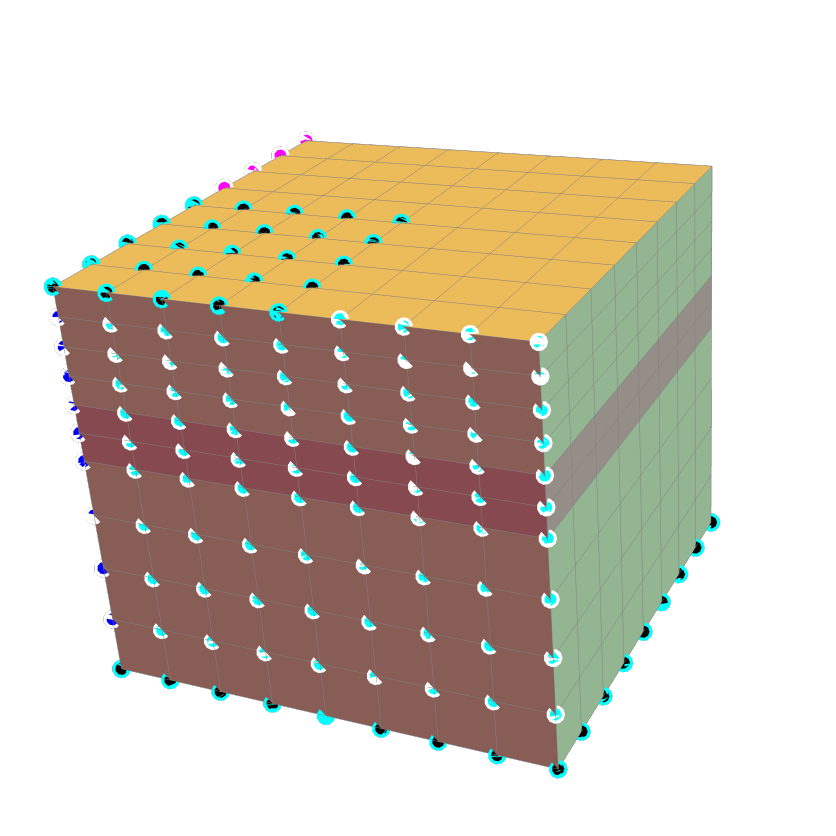}
\end{center}
\caption{System: block and a plate  with the third medium  and the loading (left) and $8\times 8 \times 10$  mesh of one quarter of the block including boundary conditions (right).}
	\label{fig:Block_3d_system}
\end{figure}

 The example of two plates that are located on top of each other with a third medium inbetween is shown in Fig. \ref{fig:Block_3d_system}.  The length, height and width of the lower block and the upper plate are $8 \times 8 \times 2$ and $8 \times 8 \times 1$, repectively. Inbetween is the third medium which has dimensions $8 \times 8 \times 0.5$.
The upper and lower block consist of the same Neo Hooke material with a modulus of compression $K=20$ and a shear modulus of $\mu=10$, furthermore the heat conductance coefficient is chosen as $k_\theta =10$ and the coefficient of thermal extension $\alpha_{t}=10^{-5}$.  The third media is modeled using $\gamma=10^{-4}$, $\beta_1=1$ and $\beta_2=10^{-2}$. The coefficient of heat conduction in the their media is $k_{TM}=10^{-1}$ with a coefficient of thermal extension $\alpha_{TM}=10^{-5}$.

The lower block is fixed in all coordinate direction at its bottom and has zero temperature at the bottom. The upper block is loaded in the red area with an extension of $4\times 4$ in the $x$-$y$ plane.The load consists of a prescribed vertical displacement $\bar u_z = -1.4$ and a temperature of $\bar T = 100$. The horizonal displacements $u_x$ and $u_y$ in the read area are fixed throughout the loading. The loading is applied in 7 steps using an adaptive load control procedure. 
 
 The system is discretized using  two different  meshes with $8\times 8 \times 10=640$ and $16 \times 16 \times 20=5.120$  H1 elements, including the third medium. This leads with 3 unknowns per node for the displacement components $u_i$, one unknown $T$ and with 3 unknowns for the regularization variables $p_i$ (only third medium) to a total of  5.643 and 40.341  unknowns for the two discretizations. 
 
Figure?\ref{fig:Block_3d_defo} illustrates the final deformation state for both mesh configurations. To enhance the visibility of the deformation in the two solid bodies, the third medium has been omitted from the visualization. The accurate enforcement of the contact constraint is clearly observable, accompanied by a noticeable lift of the unloaded portion of the upper plate.
    \begin{figure}[t!]%
	\centering
	\includegraphics[width=0.49\textwidth]{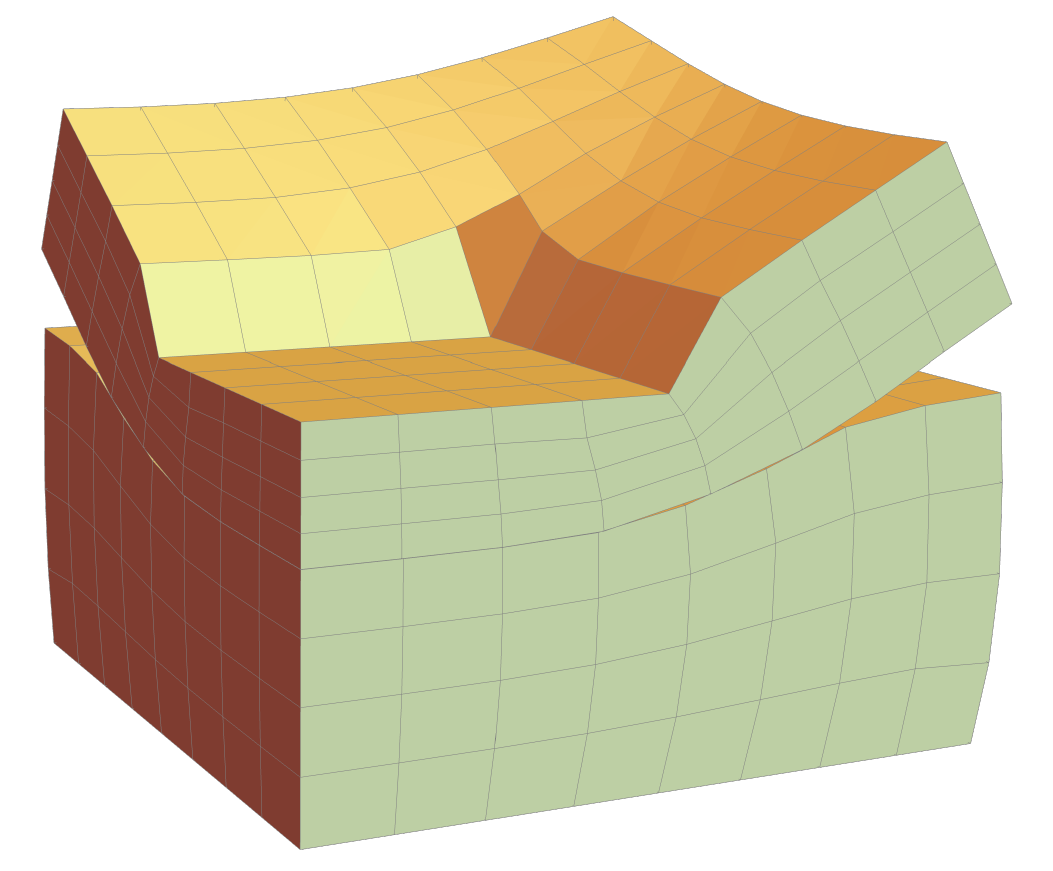}
	\includegraphics[width=0.49\textwidth]{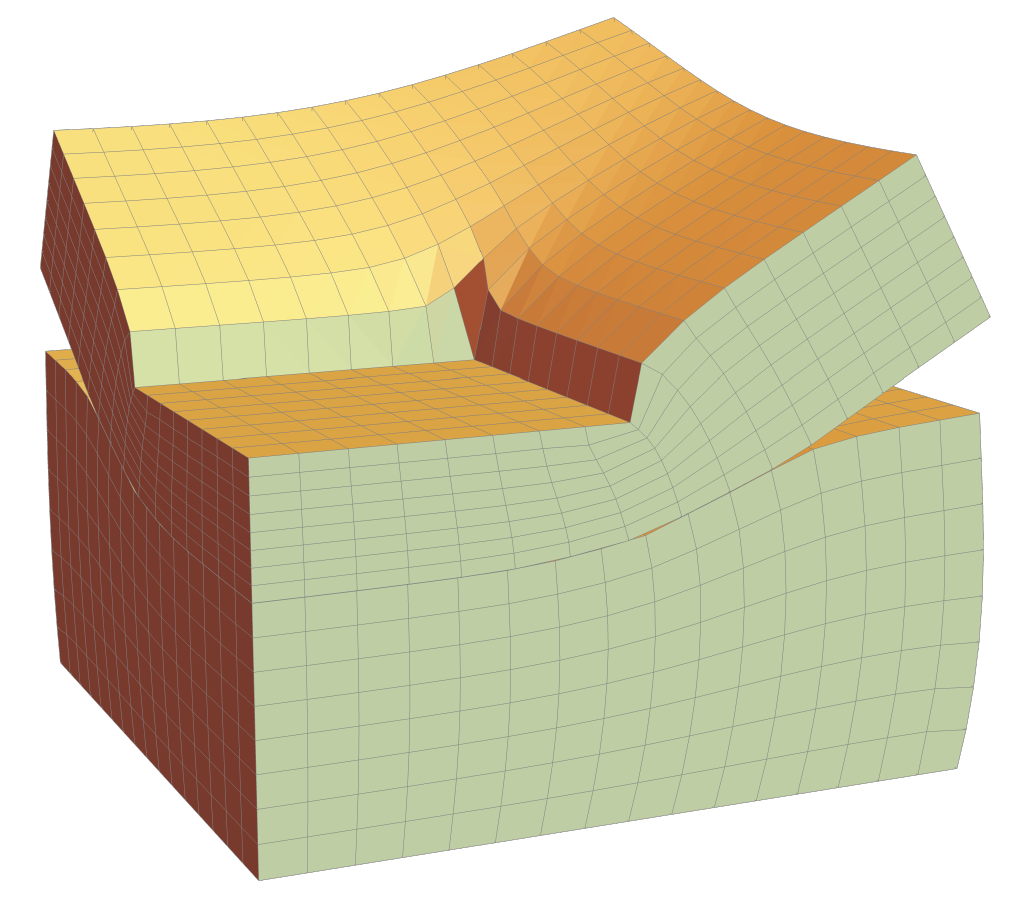}
	\caption{Deformation under a prescribed displacement for the $8\times 8 \times 10$ and $16 \times 16 \times 20 $ meshes.}
	\label{fig:Block_3d_defo}
\end{figure}%

One can observe a tangential deformation along the contact interface, which visually resembles a node-to-surface contact state. Using the $8 \times 8 \times 10$ mesh, the simulation required a total of 28 load steps and 164 iterations to converge to the final deformation. The computation with the finer mesh, as expected, required slightly more load steps (36) and iterations (231) - reflecting its increased resolution and numerical detail.

   \begin{figure}[h!]%
	\centering
	\includegraphics[width=0.49\textwidth]{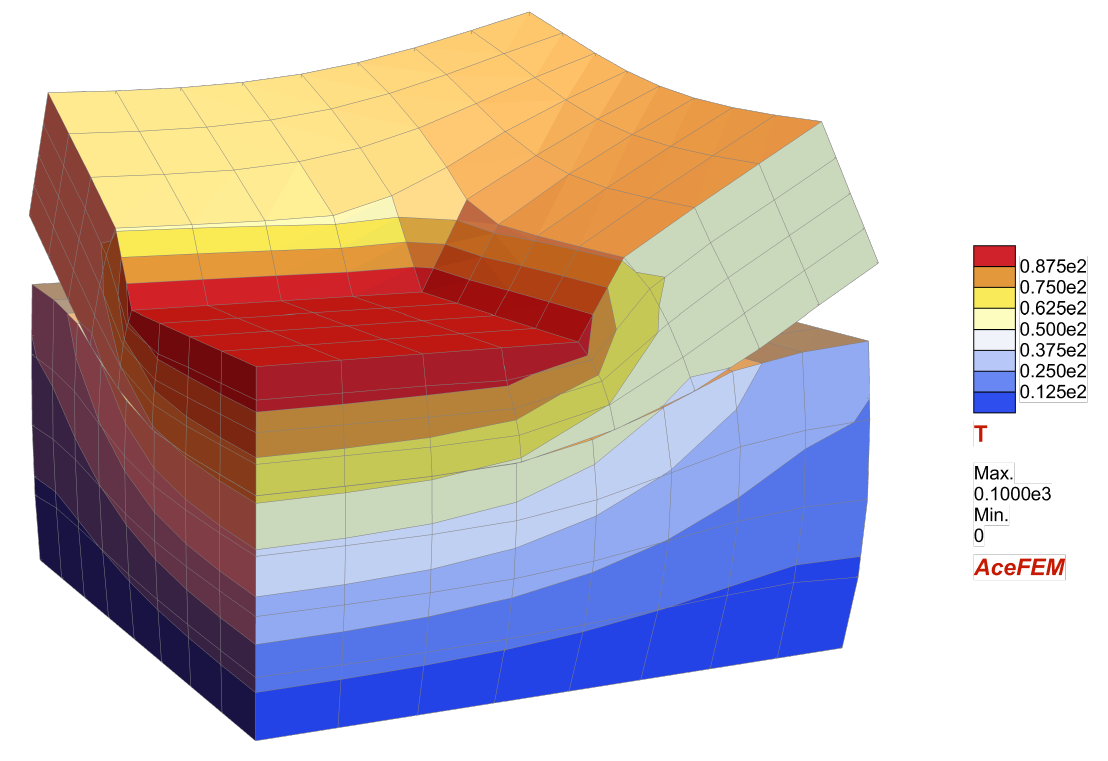}
	\includegraphics[width=0.49\textwidth]{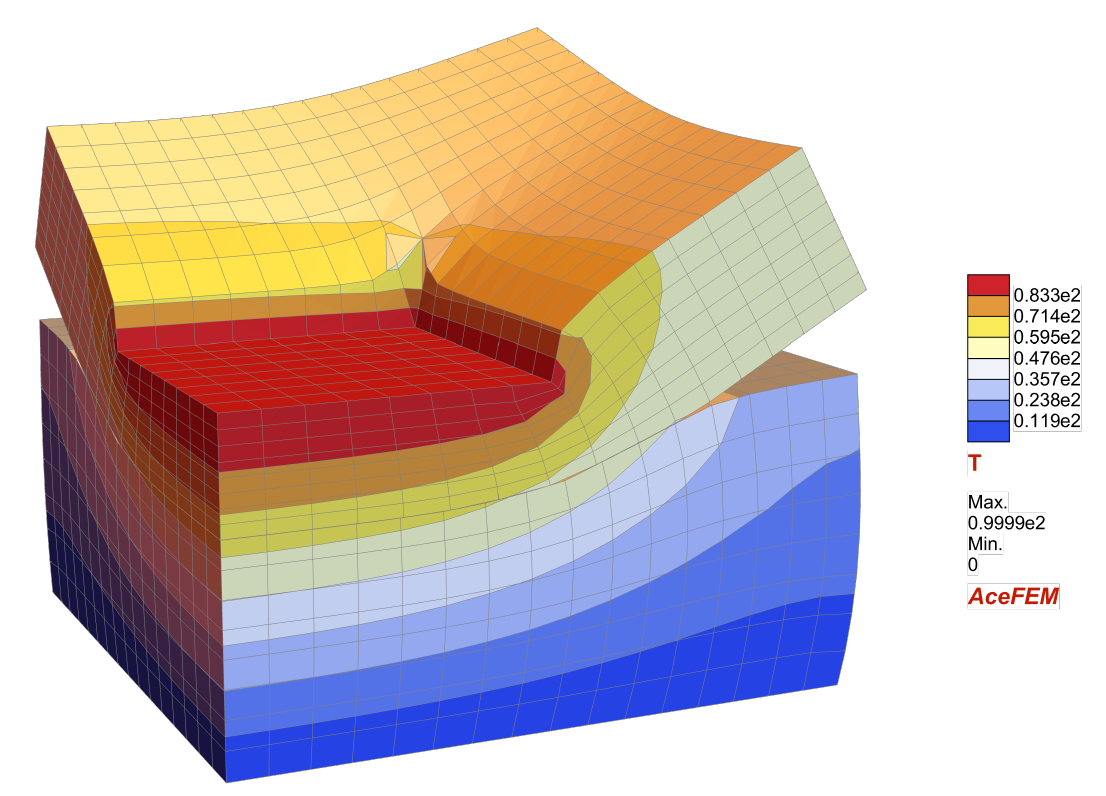}
	\caption{Temperature distribution at final loading state for meshes $8\times 8 \times 10$ and $16 \times 16 \times 20 $.}
	\label{fig:Block_3d_T+defo}
\end{figure}%
Figure \ref{fig:Block_3d_T+defo} presents the temperature distribution and confirms that temperature continuity across the contact interface is well captured - even with the coarsest mesh on the left. Notably, at the center of the block ($x = y = 0$), the temperature profile is nearly linear, whereas the remainder of the solid exhibits a visibly non-uniform temperature field.

\begin{figure}[h!]%
	\centering
	\includegraphics[width=0.49\textwidth]{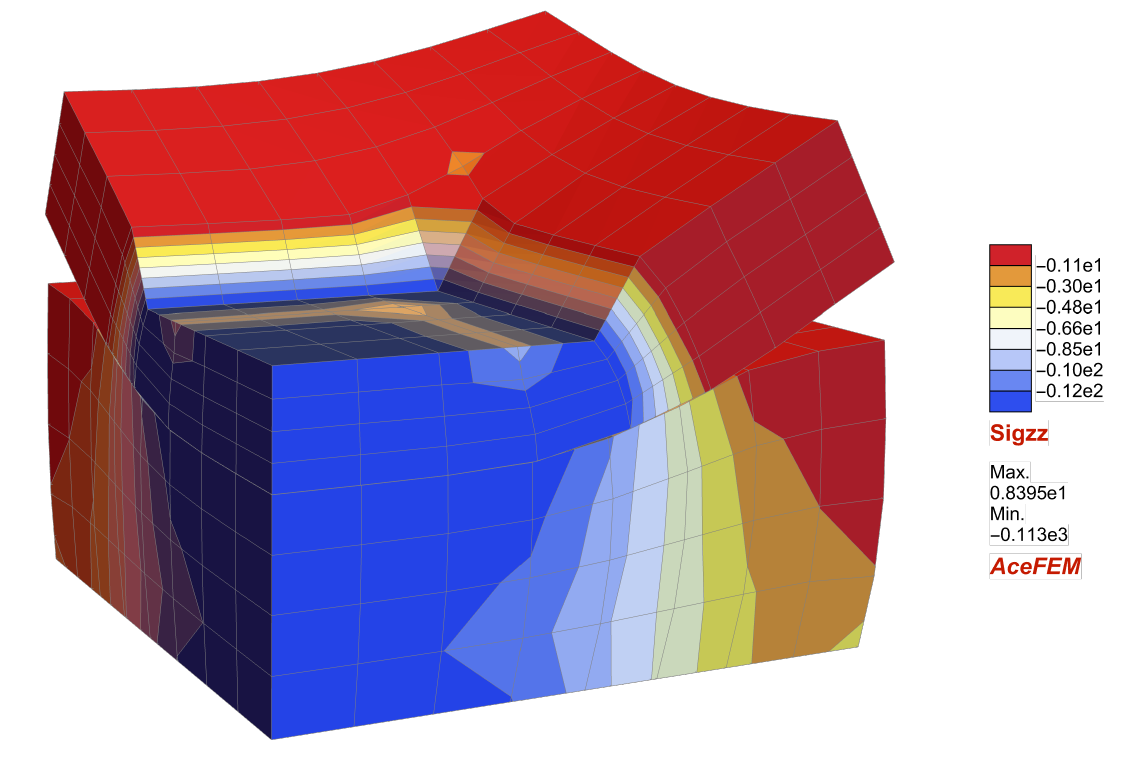}
	\includegraphics[width=0.49\textwidth]{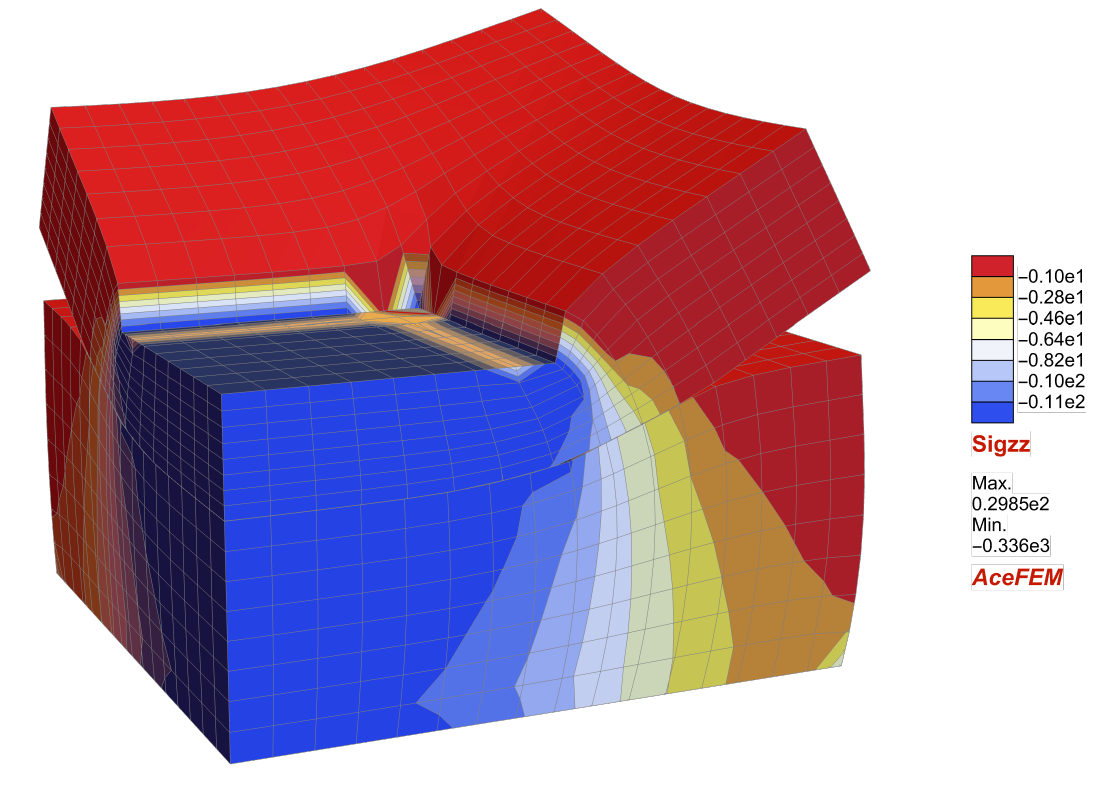}
	\caption{Cauchy stress $\sigma_{zz}$ in the deformed configuration for meshes $8\times 8 \times 10$ and $16 \times 16 \times 20 $.}
	\label{fig:Block_3d_SigZZ+defo}
\end{figure}

Figure \ref{fig:Block_3d_SigZZ+defo} shows the distribution of the Cauchy stress component $\sigma_{zz}$. On the finer mesh, higher stress magnitudes are evident, a consequence of its enhanced ability to resolve the corner singularity within the loaded region. Additionally, the stress field near the contact zone appears noticeably smoother in the results obtained with the finer mesh.

\begin{figure}[h!]%
	\centering
	\includegraphics[width=0.49\textwidth]{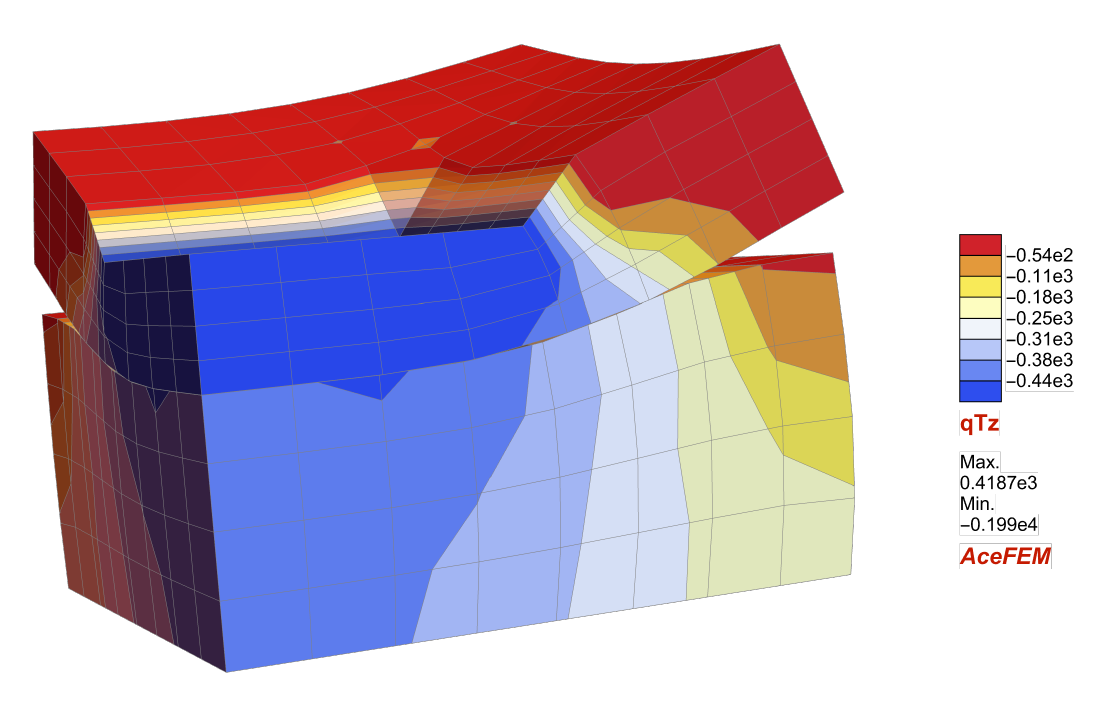}
	\includegraphics[width=0.49\textwidth]{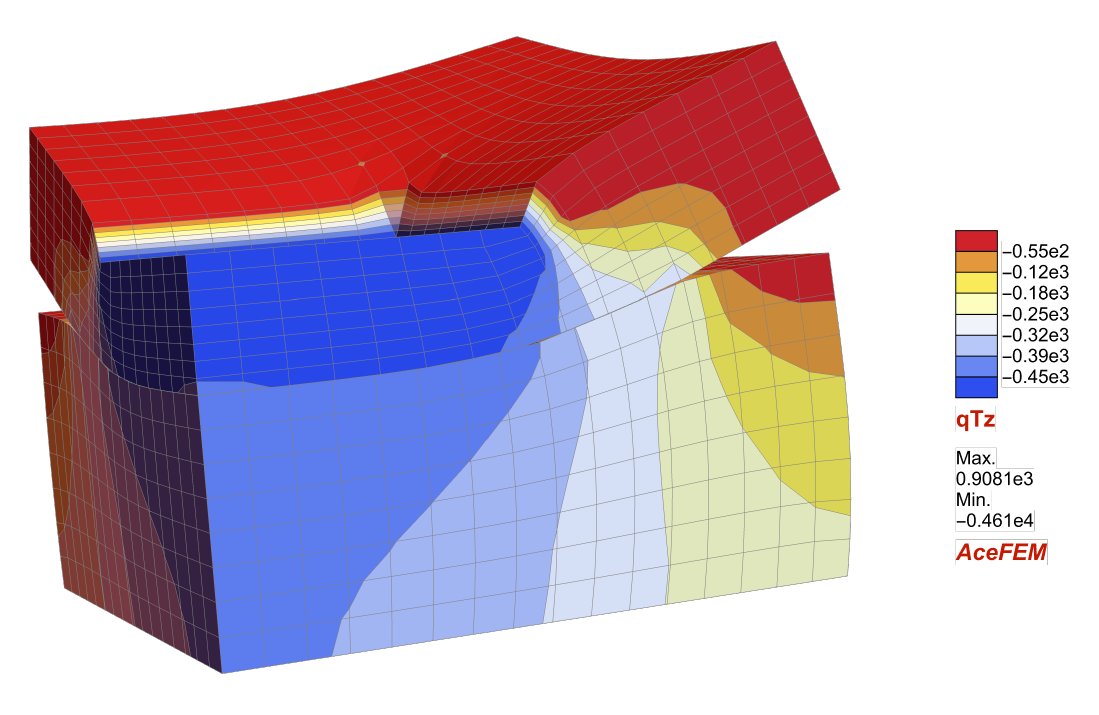}
	\caption{Component  $q_{z}$ of the heat flux in the deformed configuration for meshes $8\times 8 \times 10$ and $16 \times 16 \times 20 $.}
	\label{fig:Block_3d_qTz+defo}
\end{figure}

The heat flux component $q_z$ is shown in Fig. \ref{fig:Block_3d_qTz+defo}, demonstrating once again that both mesh resolutions reliably predict the heat flux. Moreover, the results clearly reflect the pressure-sensitive nature of thermal conduction across the interface: as contact pressure increases, so does the real contact area, resulting in higher thermal conductance and thus increased heat flux.

Finally, Fig. \ref{fig:Block_3d_defo} presents the deformed configuration of the fine mesh from the rear view. The third medium mesh remains well-controlled- without any visible bulging - demonstrating the robustness of the formulation introduced in Section \ref{sec:reduced_grad}.
    \begin{figure}[h!]%
	\centering
	\includegraphics[width=0.49\textwidth]{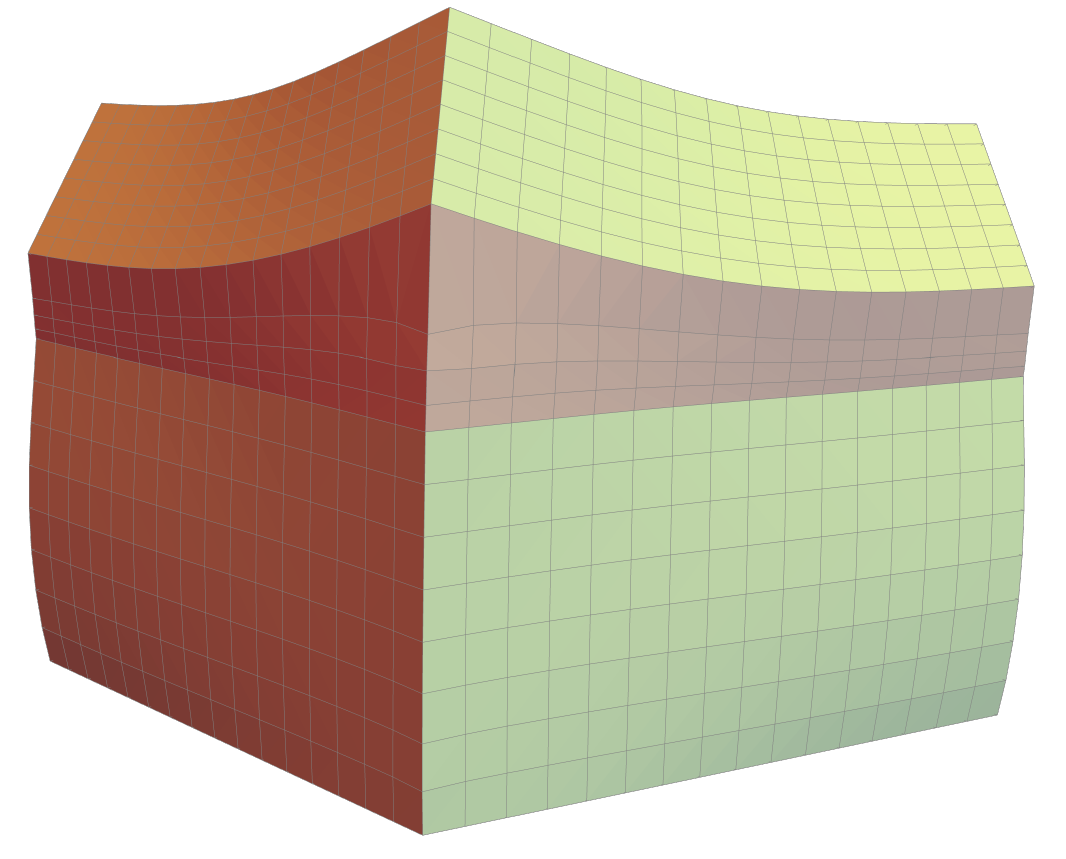}
	\includegraphics[width=0.49\textwidth]{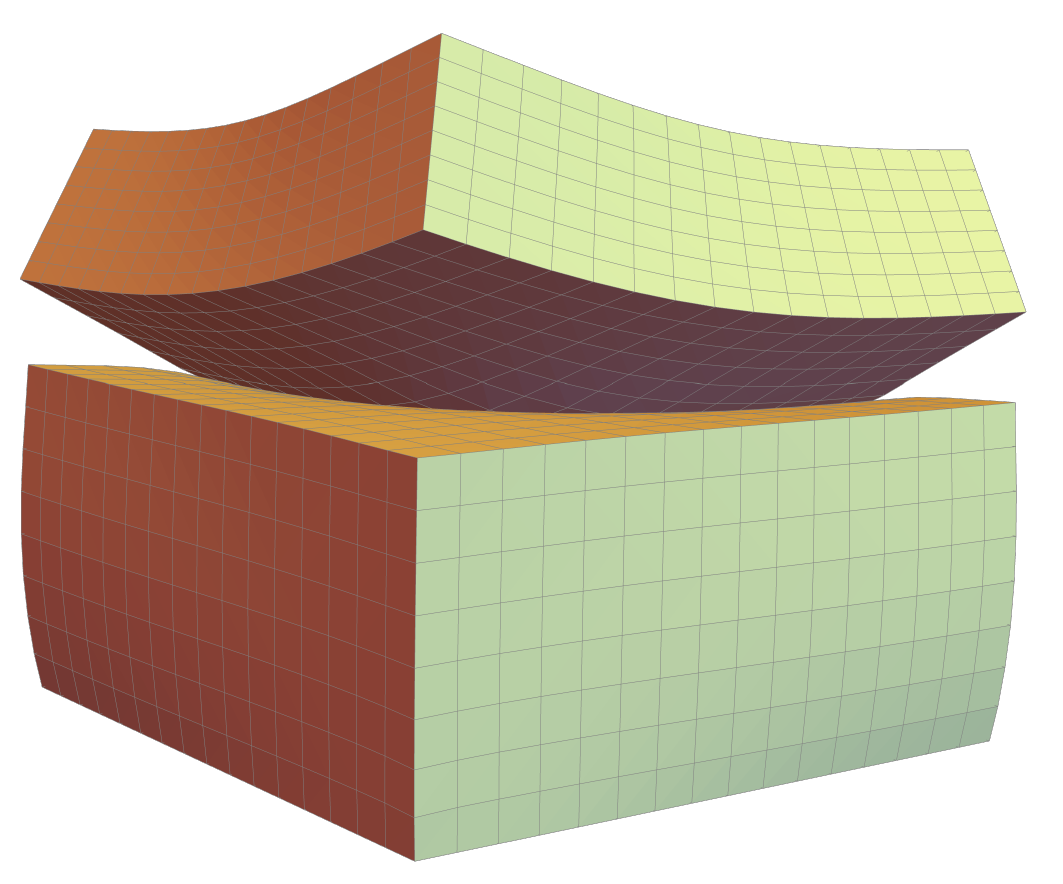}
	\caption{Deformed mesh from the back side with and without third medium.}
	\label{fig:Block_3d_defo}
\end{figure}%

A sensitivity study was conducted on the $8 \times 8$ mesh by progressively reducing the number of element layers across the height of the third medium. The results, presented in Table?\ref{tab:Block_3d_8x8}, demonstrate a clear advantage when using only a single layer of elements in the third medium for this problem type. Despite maintaining equivalent accuracy, this configuration exhibits greater computational robustness - requiring fewer load steps and iterations - which ultimately leads to significantly reduced computation times.

\begin{table}[h!]
  \begin{center}
    \caption{Number of iterations and solution times for  $n_{TH}$ layers of $8 \times 8$ H1 elements of the third medium in $z$-direction; with $\bar u_z= 1.2$ and $\gamma = 10^{-4}$.}
    \label{tab:Block_3d_8x8}
    \begin{tabular}{r|r|c|c|r|c} 
 $n_{TH}$  &  d.o.f. & load steps &iterations& solution time & gap $g_z$.\\
      \hline
1  & 5,094  & 29& 189 &4.97  s &$6.3 \cdot 10^{-3}$ \\
 2  &   5,643   & 47& 278 & $7.60$ s & $6.4  \cdot 10^{-3}$ \\
  4 & 6,741 & 63& 398 & $14.05$ s & $6.5 \cdot 10^{-3}$ 
    \end{tabular}
  \end{center}
\end{table}

\begin{figure}[h!]%
	\centering
	\includegraphics[width=0.49\textwidth]{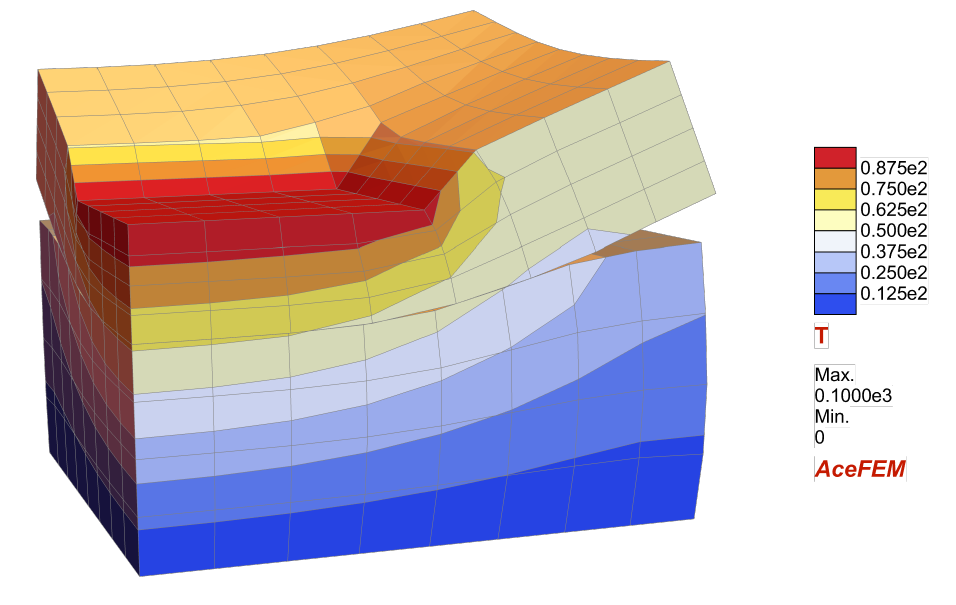}
	\includegraphics[width=0.49\textwidth]{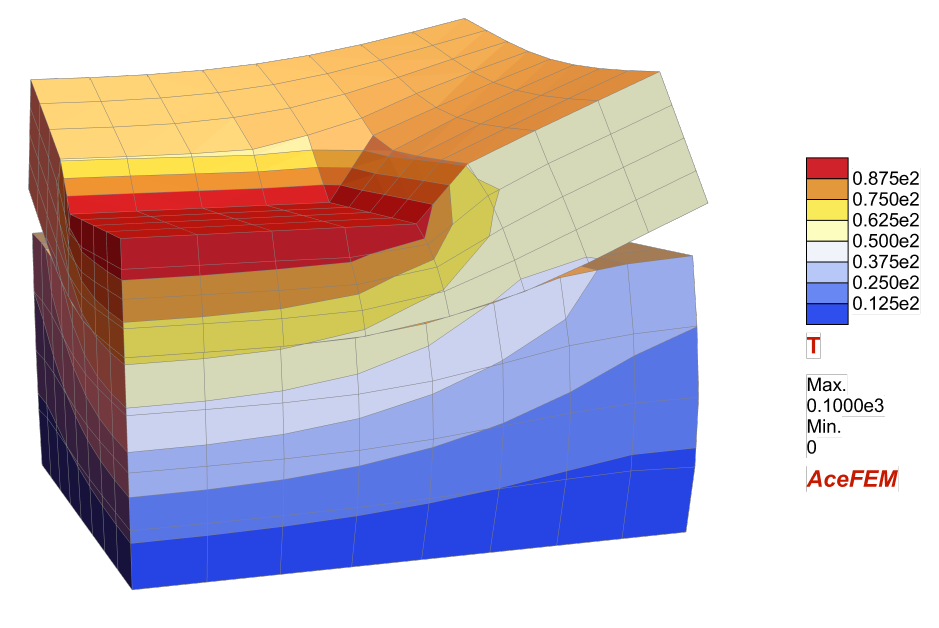}
	\caption{Temperature distribution for solutions with one and four layers of third media elements.}
	\label{fig:Block_3d_defo_LL}
\end{figure}

Figure \ref{fig:Block_3d_defo_LL} shows the temperature distributions for both the single-layer and four-layer configurations of the third medium. Notably, there is no observable difference between the two, indicating that the solution is temperature-wise insensitive to mesh layering besides large horizontal movements in the contact interface.
  
\begin{figure}[h!]%
	\centering
	\includegraphics[width=0.49\textwidth]{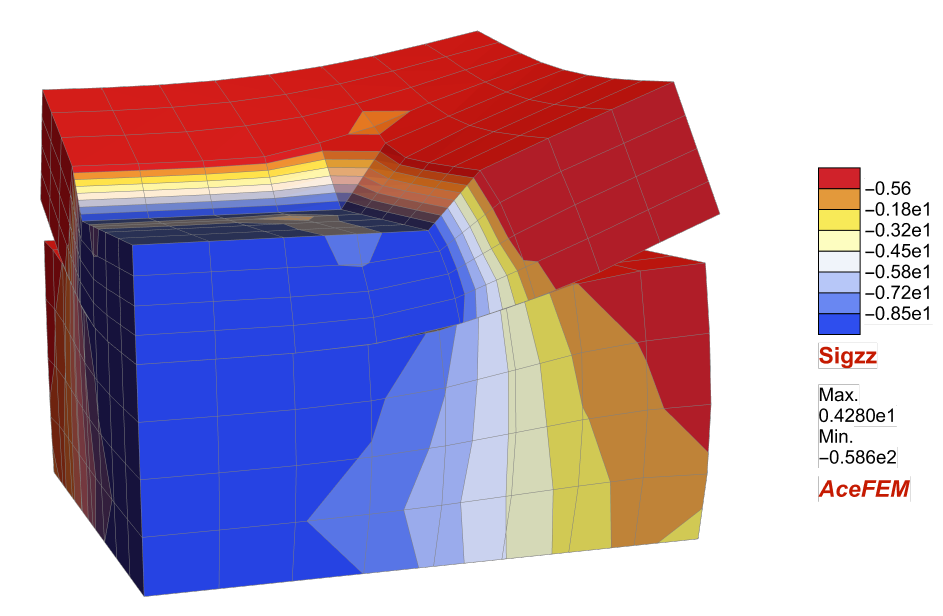}
	\includegraphics[width=0.49\textwidth]{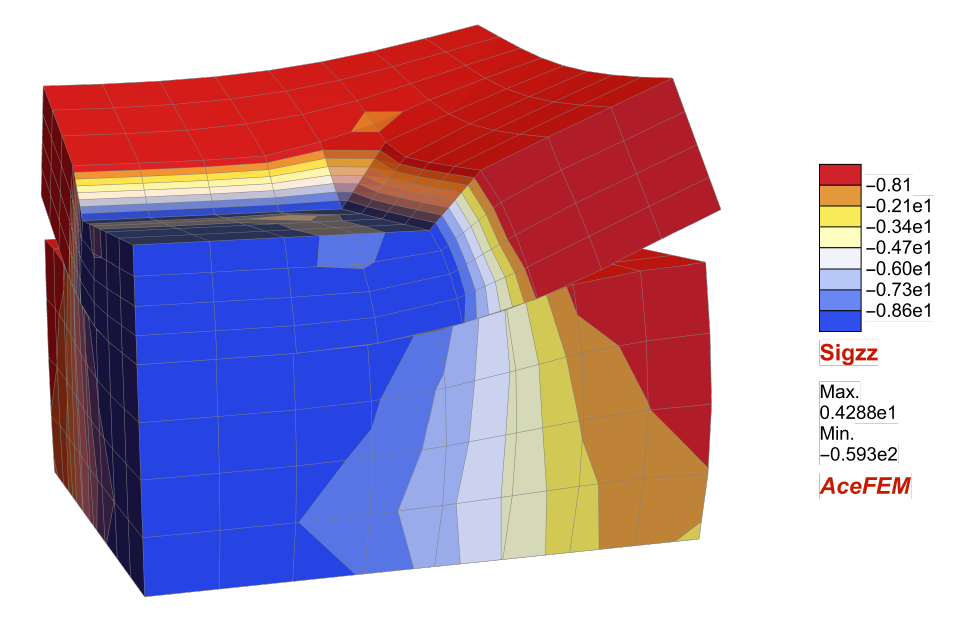}
	\caption{Distribution of the Cauchy stress component $\sigma_{zz}$ for solutions with one and four layers of third media elements.}
	\label{fig:Block_3d_Szz_LL}
\end{figure}

  Figure \ref{fig:Block_3d_Szz_LL} compares the distribution of the Cauchy stress component $\sigma_{zz}$ for configurations with one versus four layers in the third medium. The results reveal only marginal differences, reaffirming that, in simple contact scenarios, reducing the third-medium layer to a single element suffices without compromising accuracy. This simplification improves computational efficiency. However, this conclusion no longer holds in more complex setups - such as those examined in Section  \ref{sec:two_discs}, e.g., Fig. \ref{fig:Sp_ellipse_defo} - where the deformation and stress patterns become significantly more intricate.

\subsection{Comparison of the different regularizations}
This section discusses and compares for the three-dimensional case the performance of the linear third medium H1 element with 7 degrees of freedom per node $I$ - $\{\mathbf u_I\,,\theta_I\,,\mathbf p_I \}$ - when using the approximation of the regularization \eqref{eq:reg_TMC_3d} versus the quadratic H2 
elements which have generally 4 degrees of freedom per node $I$ - $\{\mathbf u_I\,,\theta_I \}$ and use the regularization \eqref{eq:W_reg_Hu_skew}. 

\begin{figure}[h!]%
	\centering
	\includegraphics[width=0.45\textwidth]{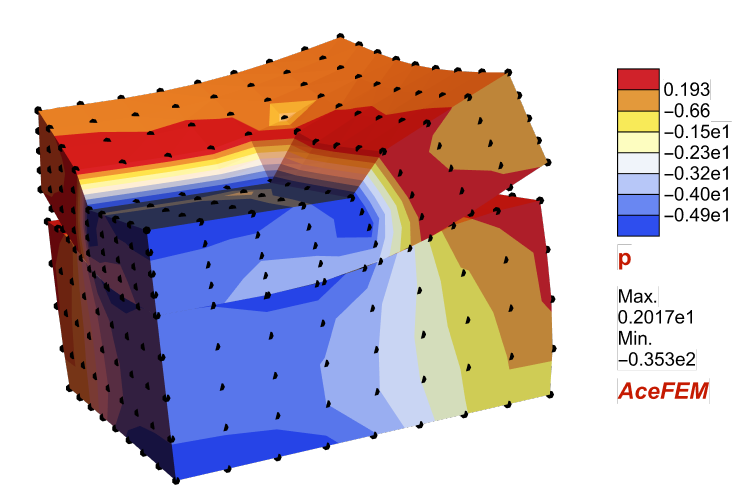}
	\includegraphics[width=0.45\textwidth]{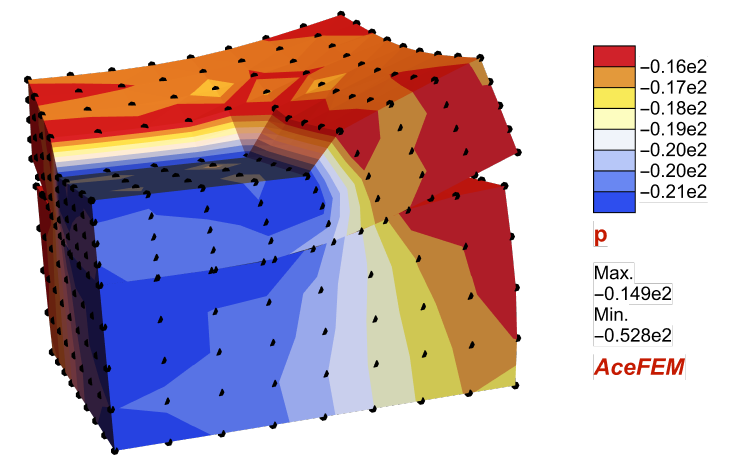}
	\caption{Distribution of the pressure $p$   for H1  and H2 elements.}
	\label{fig:Block_3d_p+defo_comparison}
\end{figure}

We use the previous example with the same data but different discretizations. The parameters for $\gamma$, $\beta_1$, $\beta_2$ and accordingly $\alpha_r$ are selected as in the last section which leads to a gap approximation of $g_z = 6.3 \cdot 10^{-4}$ for all discretizations. Pressure contours were computed using the $4\times 4$ H2 meshes which relate to a $8 \times 8$ mesh for the H1 discretization. The results are depicted in Fig. \ref{fig:Block_3d_p+defo_comparison} and illustrate that a good approximation is obtained for all discretizations,  even with these coarse meshes.

\begin{figure}[h!]%
	\centering
	\includegraphics[width=0.42\textwidth]{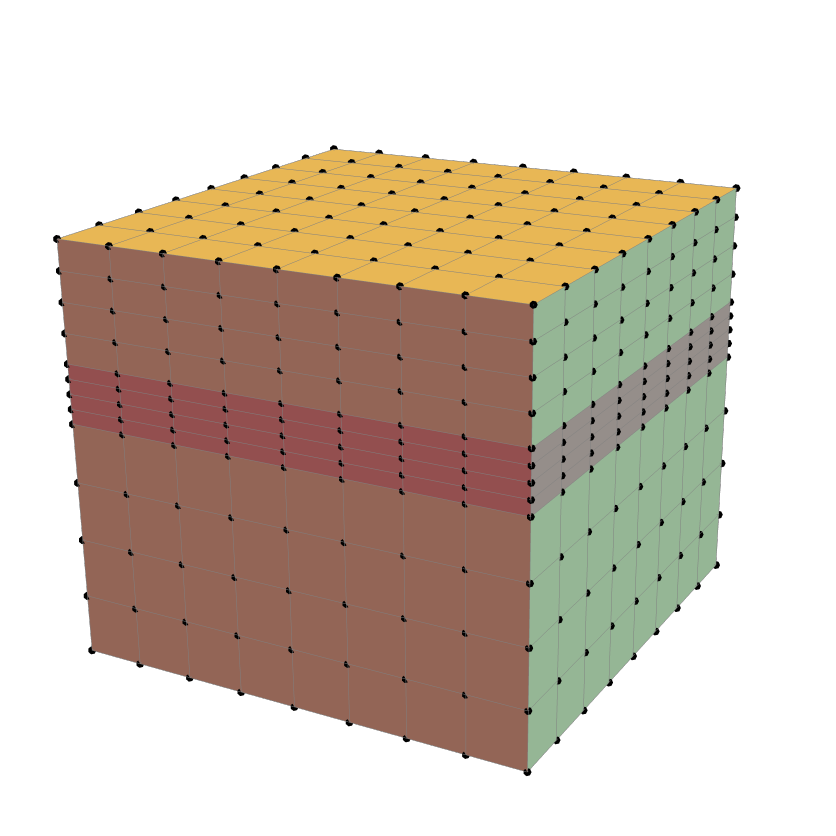}
	\includegraphics[width=0.42\textwidth]{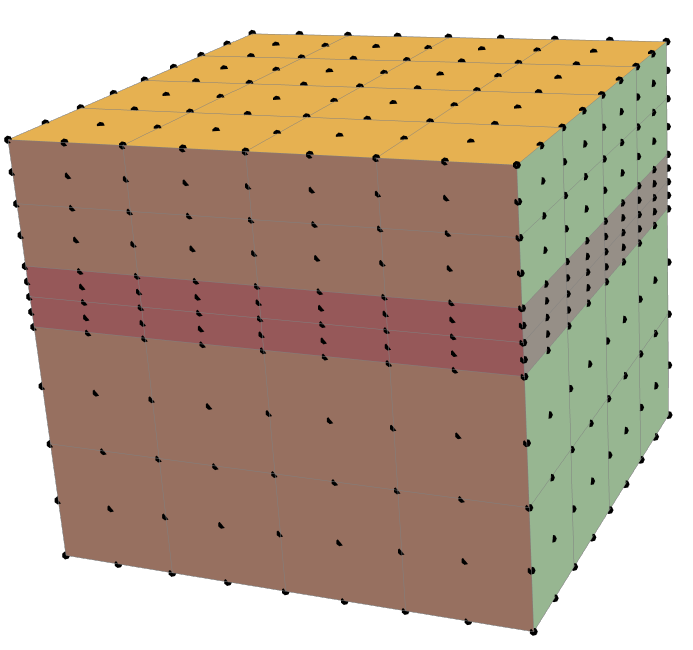}
	\caption{H1 and H2 meshes.}
	\label{fig:meshes_comparison}
\end{figure}

The idea is to compare the different discrizations in terms of  robustness  (total number of iterations), performance (solution times) and accuracy. For this purpose, meshes with the same number of nodes are employed, as shown in Fig. \ref{fig:meshes_comparison}. Based on the results of the last sections we keep $n_{TH} =1$ constant for all element formulations to ensure optimal performance.

The number of nodes  for the H1 and H2 elements is almost the same (the small difference results from the fact that there is one row of nodes more in the third medium),  
see Tab. \ref{tab:Block_3d_comparison}. Within three different discretizations the performance of the H1 and H2 
elements are compared. 

\begin{table}[b!]
  \begin{center}
    \caption{Comparison of computing times, load steps and iterations for H1 and H2 
 elements using $n_{TH}=1$ layer for the third medium.}
    \label{tab:Block_3d_comparison}
    \begin{tabular}{r|r|c|c|r|r|r} 
 Disc. (H2) & el & nodes / d.o.f. & load steps &iterations& solver time & K \& R time\\
      \hline
$4 \times 4$  & H1& 0.89 k / 3.15  k & 25& 191 & 1.60  s &0.39 s\\
 & H2  &  0.89 k / 2.97 k  & 44& 261 & $1.86$ s & 0.93 s \\
       \hline
$8 \times 8$  & H1& 5.20 k / 20.5  k & 26 & 203 & 26.7  s &2.3 s\\
 & H2  &  5.49 k / 19.9  k  & 49 & 247 & $31.1$ s & 5.3 s\\
       \hline
       $16 \times 16$  & H1& 37 k / 147  k & 27& 197 & 1070  s &15.4 s\\
 & H2  & 38 k / 144 k  & 42& 214 & $1200$ s & 40.4 s \\
    \end{tabular}
  \end{center}
\end{table}

Table \ref{tab:Block_3d_comparison} summarizes the results, demonstrating that the H1 element with a single layer of third-medium elements ($n_{TH}=1$) delivers the most robust performance across all mesh resolutions - requiring the fewest load steps. 
It is interesting to note that the solution times for the H1 and H2 discretizations are approximately the same when the number of nodes is equal. In this case, the additional degrees of freedom associated with gradient approximation in the H1 discretization appear to offset its advantage, resulting in comparable simulation times than those of the higher-bandwidth quadratic H2 discretization.

\textit{Remark:} In this study, the direct solver Pardiso (Intel MKL, \cite{ScGa04}) was employed which  is not the optimal choice for for large-scale three-dimensional problems. For instance, in the $16 \times 16$ mesh case, solving a system of approximately 144,000 equations takes around 6 seconds per solve (using a MacBook with M4 pro processor and 48 Gb main memory), illustrating the solver's limitations in scalability. While Pardiso is robust and can handle in-core and out-of-core scenarios effectively, performance degrades for three-dimensional problems significantly as problem size increases - e.g., grids approaching more than one million degrees of freedom may require several minutes per Newton iteration depending on the memory configuration. Here iterative solvers could be more efficient, but this is not in the focus of this study.

\section {Conclusion}

In this study, the solid bodies are modeled using classical two- and three-dimensional finite elements, which are not always optimal for accurately capturing bending or near-incompressible behavior. However, this simplification serves the primary aim of evaluating the performance and robustness of our new third-medium contact formulation - particularly its ability to accommodate large deformations and dynamic contact transitions. As shown by \cite{WrKoJu25}, the newly proposed regularization permits the use of simple linear finite elements for the third medium while maintaining accuracy comparable to more complex formulations. For applications requiring enhanced fidelity, one could instead employ advanced element types to address locking issues - as in \cite{KoSoWr10} enhanced-strain brick elements - or adopt a hybrid mesh approach combining quadratic elements for the solids with linear elements for the third medium. We intend to explore these strategies in a forthcoming publication.

Moreover, future work should focus on improving the computational treatment of the auxiliary field variables $p_i$ and on extending the framework to frictional contact scenarios, thereby incorporating frictional heating - a key feature for industrially relevant thermo-mechanical applications.

\section{Acknowledgement}
 P. Wriggers was  funded by the Deutsche Forschungsgemeinschaft (DFG, German Research Foundation) under the DFG research unit FOR 5250 ''Mechanism-based characterization and modeling of permanent and bioresorbable implants with tailored functionality based on innovative in vivo, in vitro and in silico methods'' (project number: 449916462).

\clearpage
\bibliography{nlbuch,pw,c-book}
\bibliographystyle{cont_book}
\end{document}

%% file: BH_DEFS.tex
\usepackage{amsmath}
\usepackage{ntheorem}

\usepackage{amssymb}
\usepackage{amsfonts}
\usepackage[]{algorithm2e}

\SetKw{KwAnd}{{\bf and}}
\SetKw{KwOr}{{\bf or}}
\SetKw{KwNot}{{\bf not}}
\SetKw{KwFalse}{{\bf false}}
\SetKw{KwTrue}{{\bf true}}
\SetKw{KwPrint}{{\bf print}}
\SetKw{KwStop}{{\bf stop}}

\SetKwComment{KwADBForm}{}{}
\theoremstyle{plain}
\theoremstyle{empty}


%% file: PWc_DEFS.tex
\def\pb #1{\mbox{\boldmath{$#1$}}}
\def\pbd #1 {\dot{\pb #1}}
\def\pbh #1 {\hat {\pb #1}}
\def\pbb #1 {\bar{\pb #1}}

 \DeclareSymbolFont{OpenBBM}{U}{bbm}{m}{n}
\DeclareSymbolFont{OpenBBMbold}{U}{bbm}{b}{n}
 \DeclareSymbolFontAlphabet{\mathbbA}{OpenBBM}

\newcommand {\be}{\begin{equation}}
\newcommand {\ee}{\end{equation}}

\def\fpbox #1 {\fbox{$\displaystyle #1 $}}
\def\beq {\begin{equation}}
\def\bqa {\begin{eqnarray}}
\def\ba {\begin{array}}
\def\bqas {\begin{eqnarray*}}
\def\btb {\begin{table}}
\def\bpi {\begin{picture}}
\def\bit {\begin{itemize}}
\def\ben {\begin{enumerate}}
\def\eeq {\end{equation}}
\def\eqa {\end{eqnarray}}
\def\ea {\end{array}}
\def\eqas{\end{eqnarray*}}
\def\etb {\end{table}}
\def\epi {\end{picture}}
\def\eit {\end{itemize}}
\def\een {\end{enumerate}}
\def\Reals{ {\hbox{$\mathpalette{}{\textrm I\kern-.18em R}$}}}
\def\Edom{ {\hbox{$\mathpalette{}{\textrm I\kern-.18em E}$}}}
\def\Eone{ {\hbox{$\mathpalette{}{\\textrm I\kern-.18em I}$}}}
\def\Lten{ {\hbox{$\mathpalette{}{\\textrm I\kern-.18em L}$}}}
\def\Cten{ {\hbox{$\mathpalette{}{\\textrm l\kern-.48em C}$}}}
\def\Hten{ {\hbox{$\mathpalette{}{\\textrm l\kern-.48em H}$}}}
\def\cten{ {\hbox{$\mathpalette{}{\\textrm c\kern-.28em c}$}}}
\def\aten{ {\hbox{$\mathpalette{}{\\textrm a\kern-.36em a}$}}}
\def\Proj{ {\hbox{$\mathpalette{}{\\textrm I\kern-.18em P}$}}}
\catcode`\@=11\def\endbox{
        \leavevmode\kern-.25em\raise-.15em
        \vbox{\hrule height 0.8truept
        \hbox{\vrule width  0.8truept
        \vbox{\kern 4truept
        \hbox{\kern 1.5truept
        {}\kern 2.5truept}
          \kern 3.0truept}
        \vrule width  0.8truept}
        \hrule height 0.8truept}
        }

\catcode`\@=11
\newdimen\ex@
\def\boxed#1{\setbox\z@\hbox{$\displaystyle{#1}$}\hbox{\lower.4\ex@
 \hbox{\lower3\ex@\hbox{\lower\dp\z@\hbox{\vbox{\hrule height.4\ex@
 \hbox{\vrule width.4\ex@\hskip3\ex@\vbox{\vskip3\ex@\box\z@\vskip3\ex@}%
 \hskip3\ex@\vrule width.4\ex@}\hrule height.4\ex@}}}}}}
\catcode`\@=\active